\RequirePackage{fix-cm}

\documentclass[smallextended]{svjour3}
\smartqed

\usepackage[T1]{fontenc}
\usepackage{cite}
\usepackage{algorithmic}
\usepackage{graphicx}
\usepackage{textcomp}
\usepackage{xcolor}
\usepackage{booktabs}
\usepackage{footmisc}
\usepackage{longtable}

\usepackage{microtype}
\usepackage{siunitx}
\usepackage{minibox,soul}
\usepackage{xspace}
\usepackage[inline]{enumitem}
\usepackage[xspace]{ellipsis} % must be after hyperref
\usepackage[margin= 1.0in]{geometry}
\usepackage[section]{placeins}

\usepackage{rotating}
\usepackage{longtable}
\usepackage{multirow}
\usepackage{soul}
\usepackage{hyperref}
\usepackage{framed}
\usepackage{lscape}
\usepackage{amsmath,scalerel}

\usepackage{pifont}
\usepackage{selinput}
\usepackage{rotating}
\usepackage{capt-of}
\usepackage{afterpage}
\usepackage{pdflscape}
\usepackage{inputenc}
\usepackage{amssymb}

\definecolor{Mygray}{rgb}{0.90,0.90,0.90}

\UseRawInputEncoding

\begin{document}

\title{Free Open Source Communities Sustainability: Does It Make a Difference in Software Quality?}
%\title{Unveiling the symbiosis: psychological safety, and the quest for software quality in agile teams}

\author{Adam Alami \and Ra{\'u}l Pardo \and Johan Lin\r{a}ker}

\institute{A. Alami \at
              Aalborg University \\
              Department of Computer Science \\
              Selma Lagerl\"ofs Vej 300  \\
              9220 Aalborg,  Denmark \\
              \email{adal@cs.aau.dk}
              \and
           R. Pardo\at
              IT University of Copenhagen\\
              Denmark\\
              \email{raup@itu.dk}
              \and
           J. Linaker\at
              RISE Research Institutes of Sweden \\
              Sweden \\
              \email{johan.linaker@ri.se}
}

\date{Received: date / Accepted: date}
% The correct dates will be entered by the editor

\maketitle

\begin{abstract}

\textbf{Context:} Free and Open Source Software (FOSS) communities' ability to stay viable and productive over time is pivotal for society as they maintain the building blocks that digital infrastructure, products, and services depend on. Sustainability may, however, be characterized from multiple aspects, and less is known how these aspects interplay and impact community outputs, and software quality specifically.

\textbf{Objective:} This study, therefore, aims to empirically explore how the different aspects of FOSS sustainability impact software quality.

\textbf{Method:} 16 sustainability metrics across four categories were sampled and applied to a set of 217 OSS projects sourced from the Apache Software Foundation Incubator program. The impact of a decline in the sustainability metrics was analyzed against eight software quality metrics using Bayesian data analysis, which incorporates probability distributions to represent the regression coefficients and intercepts.

\textbf{Results:} Findings suggest that selected sustainability metrics do not significantly affect defect density or code coverage. However, a positive impact of community age was observed on specific code quality metrics, such as risk complexity, number of very large files, and code duplication percentage. Interestingly, findings show that even when communities are experiencing sustainability, certain code quality metrics are negatively impacted.

\textbf{Conclusion:} Findings imply that code quality practices are not consistently linked to sustainability, and defect management and prevention may be prioritized over the former. Results suggest that growth, resulting in a more complex and large codebase, combined with a probable lack of understanding of code quality standards, may explain the degradation in certain aspects of code quality.


\end{abstract}

\section{Introduction}\label{sec:introduction}

Free and Open Source Software (FOSS) has become an essential component of our contemporary digital infrastructure \cite{crowston2008free}. It is utilized in a wide range of applications, from web servers to data analysis tools, and has allowed worldwide innovation and cooperation. Still, practitioners and academics alike continue to worry about FOSS's sustainability \cite{fang2009understanding}. The ability of open source actors to draw in resources like developers or value like knowledge has traditionally been the primary focus of sustainability concerns \cite{chengalur2010sustainability}. As recent global cybersecurity issues like log4j \cite{Log4j}, a well-known and extensively used FOSS project that is scarcely maintained by a group of just four unpaid developers, have revealed, the repercussions of a prominent FOSS community becoming unsustainable may be disastrous \cite{stuanciulescu2022code}. Similar incidents, such as OpenSSL's ``Heartbleed,'' have been reported in the past, compromising the security of a major amount of protected Internet communication \cite{walden2020impact}. Despite its widespread use, OpenSSL received little attention in terms of maintenance due to a shortage of labor \cite{champion2021underproduction}. 

Chengalur-Smith et al. \cite{chengalur2010sustainability} define sustainability as ``the ability of an organism or an ecosystem to maintain its activity and productivity over time.'' In the same vein, Lin{\aa}ker et al. \cite{linaaker2022characterize} define FOSS health as ``project's capability to stay viable and maintained over time without interruption or weakening.'' It should be noted that the terms ``health'' and ``sustainability'' are used interchangeably in the literature \cite{linaaker2022characterize}. In this study, we adopted ``sustainability'' to remain consistent with recent work, for example, \cite{yin2021sustainability,stuanciulescu2022code}. 

FOSS sustainability is critical for these digital platforms to continue encouraging accessibility by making technology accessible to a larger audience and allowing people and organizations to freely use, modify, and distribute software regardless of their economic or geographic situation. However, the FOSS sustainability impact on the community's product is less known. While recent FOSS sustainability research has focused on forecasting sustainability \cite{yin2021sustainability}, reasons for project failure \cite{coelho2017modern}, community health indicators \cite{xia2022predicting,manikas2013reviewing,linaaker2022characterize}, project popularity \cite{han2019characterization,borges2016understanding}, developers and users attraction to the community \cite{chengalur2010sustainability}, little work has been devoted to the implications of FOSS sustainability and its various aspects on community outputs. Software quality (SWQ) is deemed an important aspect of these outputs \cite{vasilescu2015quality}. 

Extant work has investigated specific aspects of FOSS sustainability, such as developer's turnover \cite{foucault2015impact}, effort distributions \cite{wang2020unveiling}, or bug-fixing \cite{ghapanchi2015predicting}. For example, Foucault et al. investigated the relationship between developer turnover and modules' quality \cite{foucault2015impact}, Wang et al. investigated the effort distributions of elite developers on productivity and quality \cite{wang2020unveiling}, and Ghapanchi investigated the capacity of a FOSS project to fix bugs, add new features, and manage releases in relation to its ability to remain sustainable \cite{ghapanchi2015predicting}. However, these studies offer fragmented insights into the relationship between various dimensions of sustainability and their collective impact on SWQ. In addition, previous studies typically focused on individual sustainability indicators, one at a time. 

The research by Stuanciulescu et al. indicated that the chance of graduating from the Apache Software Foundation incubator (ASFI) is reliant on code quality, among other factors \cite{stuanciulescu2022code}. The study's approach, however, represents a limitation in capturing the complex nature of FOSS sustainability by using graduation or retirement as a proxy for sustainability, as this may be a consequence of a myriad of external factors, such as life-cycle events, the subjective criteria of the community-driven process of graduation (e.g., voting, board approval, etc.), and the emergence of competing projects. 

To overcome these limitations, we employed the comprehensive framework developed by Lin{\aa}ker et al. \cite{linaaker2022characterize} for assessing FOSS sustainability. The framework proposes, based on a survey of extant work, a diverse array of sustainability indicators, including communication dynamics, popularity trends, stability measures, and technical activity. The framework acknowledges that sustainability is influenced by several interconnected factors, instead of being a mere product of project graduation or retirement \cite{stuanciulescu2022code}, or a single sustainability indicator, e.g., turnover \cite{foucault2015impact}, or elite developers \cite{wang2020unveiling}. Lin{\aa}ker et al. \cite{linaaker2022characterize} framework allows for a nuanced investigation of the interplay between FOSS sustainability and software quality. By assessing sustainability through multiple lenses and indicators, we can discern how different aspects, like community engagement or the consistency of technical contributions, impact the quality of software produced. For example, active and consistent community engagement may contribute to better bug fixing, thereby resulting in better quality. To meet these objectives, we ask:

\medskip 

\noindent \textbf{\emph{RQ:}} How does the decline in the sustainability of FOSS communities affect their software quality? 

\medskip

Aligning with St{\u{a}}nciulescu, et al.~\cite{stuanciulescu2022code}, we selected projects from the ASFI program. Our data covered projects' activities in the period from March 2009 to April 2023. These projects are not only diverse in scope but also at different stages of development, which makes our dataset contextually rich. Our dataset includes 217 projects after applying our selection criteria. We used Git and Perceval to clone repositories and collect relevant repository data. We adopted Bayesian statistics for data analysis. Our model includes probability distributions for the regression coefficients and intercepts.

We found no statistical evidence of positive or negative impacts of our selected sustainability metrics on defect density, code coverage, or code duplication percentage. However, community age impacts some code quality metrics positively, namely very high risk complexity, the number of very large files, the very large function size count, and the most complex function. Interestingly, our analysis shows that some sustainability metrics within the themes of stability (e.g., size) and communication (e.g., response time) can still have a negative impact on some code quality metrics, even when communities may be considered sustainable. This implies that code quality practices are not necessarily linked to sustainability. In contrast, St{\u{a}}nciulescu, et al.'s work suggested differing code quality outcomes for sustainable and unsustainable projects \cite{stuanciulescu2022code}, such as the impact of file size and the presence of certain code characteristics. 

Our study takes a different approach than St{\u{a}}nciulescu, et al.'s study; we employed sustainability metrics drawn from the literature instead of using the graduation-retirement dichotomy as the yardstick for sustainability. Our approach assumes that the decisions made for graduation and retirement do not necessarily reflect the project's sustainability. This assumption is grounded in the fact that ASFI projects graduation and retirement decisions are influenced by several factors that do not always fit with what has been reported in the literature as sustainability indicators. Importantly, we do not claim that our approach is superior to the one adopted by St{\u{a}}nciulescu, et al., but rather, we view it as a legitimate and complementary avenue of investigation. In addition, while St{\u{a}}nciulescu, et al. focused on code quality \cite{stuanciulescu2022code}, we extended the outcome variable to cover both code quality and defect density, representing software quality. While both approaches have their merits, the misalignment in the findings underscores the complex relationship between FOSS sustainability and SWQ.

Even though we found no statistical evidence to show the impact of FOSS sustainability on quality, our study brings a novel contribution to the understanding of sustainability and quality in FOSS communities. Our study moved beyond single indicators and offers a comprehensive assessment of sustainability.

We contribute with practical implications for FOSS communities. For example, the analysis of our sample data suggests that even when showing signs of sustainability, communities show subpar code quality practices. This implies that communities are not attentive to some of their code characteristics, like the number of very large files, the very large function size count, and the most complex function. FOSS communities should actively promote awareness of and adherence to code quality metrics among their contributors. This can be achieved through the promotion of awareness of these code quality standards by using documentation, guidelines, and regular hackathons focusing on code quality practices. ASFI mentoring programs should also promote adhering to code quality standards.

The remainder of our paper is organized as follows: In Sect. \ref{sec:related}, we summarize earlier work on the impact of FOSS sustainability on software quality, and in Sect. \ref{sec:sustainability}, we explain how we measure FOSS sustainability and software quality in Sect. \ref{sec:swq}. In Sect. \ref{sec:hypotheses}, we present our hypotheses. In Sect. \ref{sec:methods}, we describe our research methods. Section \ref{sec:findings} is devoted to the findings of our study and possible implications of our work in Sect. \ref{sec:discussion}. Finally, before ending in Sect. \ref{sec:conclusion}, we highlight risks to the validity of our findings in Sect. \ref{sec:validity}.

\section{Related Work}\label{sec:related}

Related work includes research that looked at the relationship between FOSS sustainability or its indicators (for example, developer turnover) and software quality or some of its metrics (for example, defect fix rate). Although researchers are interested in the subject, their approach to sustainability is not holistic. Related work focuses on specific elements or features of FOSS sustainability (for example, elite developers) and does not stress the overall influence of sustainability. We identified four studies, i.e., \cite{stuanciulescu2022code,ghapanchi2015predicting,foucault2015impact,wang2020unveiling} related to ours.

St{\u{a}}nciulescu, et al. \cite{stuanciulescu2022code} postulated that there is a relationship between project graduation and code, process, and quality measures of software in ASFI projects, motivated by earlier work on FOSS success, health, and sustainability, and driven by contingency theory. While retired projects had somewhat greater cyclomatic complexity when corrected for project size, their retirement does not seem to be connected with defects, complexity, or technical debt, according to the authors. The authors discovered that both significant and small contributions help to increase the viability of ASFI projects. Among process and quality characteristics, file size, function size, and function complexity seem to be the most unfavorable, indicating that keeping workflow simple and brief is important for sustainability. The authors recommend that FOSS maintainers assist in building a positive feedback loop for all contributors, giving them greater ownership and responsibility while also allowing them to pick which features to work on and contribute to the project's long-term goal \cite{stuanciulescu2022code}. This study's used ASFI projects for retirement and graduation as a proxy for sustainability. Our work expands on this by using the Lin{\aa}ker et al. \cite{linaaker2022characterize} framework to zoom closer into various sustainability indicators reported in the literature and how they individually influence quality.

Ghapanchi \cite{ghapanchi2015predicting} examined how features of FOSS projects affect their sustainability. The results show that the rate at which defects are fixed, the rate at which new features are added, and the speed with which these features are added may all help FOSS adopters forecast the longevity of a project. FOSS projects that prioritize their users' needs in terms of software features and functionalities are more likely to be able to sustain their development activities in the subsequent period, as the rate at which these are improved has a positive effect on development sustainability. This suggests that, particularly in the beginning phases of a project's development, the quantitative elements of defect removal, functionality enhancement, and software release play a vital role in the project's sustainability \cite{ghapanchi2015predicting}. While Ghapanchi's work identified predictors of sustainability, we examined sustainability with different dimensions, using several themes, e.g., communication and stability, and their underlying indicators and their impact on SWQ.

To investigate the impact of turnover on software quality, Foucault et al. \cite{foucault2015impact} examined the source code repositories of five open-source projects. They find that although developers quitting the project had a negative influence on quality, new members had no effect. They examined bug-related data and performed Spearman correlation tests to determine the relationship between each turnover metric and the density of bug fixes per module. They discovered that whereas external turnover (leaving the community) has a negative influence on module quality, internal turnover (moving to another module within the community) has essentially no effect \cite{foucault2015impact}. In the context of ASFI projects, we found no statistical evidence of positive or negative impacts of defect density, code coverage, medium risk complexity, or code duplication percentage. However, even when projects are experiencing low turnover, certain elements of code quality (e.g., very risk complexity and the number of very large files) deteriorate.

Several studies, for example,\cite{jensen2007role,crowston2005social}, have shown that among hundreds of contributors, only a small group of exceptional coders (elite or core developers) contribute the majority of code and manage project growth. In Mockus et al.'s \cite{mockus2002two} analysis of the Apache community projects, for example, they discovered that the top 15 contributors (out of 388 overall) contributed more than 83\% of change requests and 66\% of issue reports. Understanding elite developers is crucial for analyzing the community's health and sustainability \cite{wang2020unveiling}. Wang et al. \cite{wang2020unveiling} performed empirical research utilizing data from 20 large open source projects to investigate the links between the effort distributions of elite developers and project productivity and quality. The results illustrate a quandary that elite developers often confront in their projects: they need to spend more time on nontechnical duties, yet their projects will suffer some productivity and quality loss as a result. The effort distributions of elite developers exhibit substantial connections with project productivity and quality results. When they devote more time and effort to communicative and supporting tasks, the productivity of a project (as measured by the number of new commits and bug cycle time in each project-month) is likely to suffer. Furthermore, the quality of a project (as evaluated by the number of new defects discovered in each project month) is inversely related to its actions on organizational and supporting tasks. Another metric, the bug fix rate in each project month, is positively connected with supporting actions and may rise as elite developers devote more time and effort to support. By doing nontechnical tasks to assist other community developers and users, the elites may have less time to focus on code. Meanwhile, the project may get a significant number of recommended modifications and patches from non-elite contributors, but their work may be more bug-ridden (quality loss) \cite{wang2020unveiling}. Our study provides a broader perspective by using several sustainability themes (communication, popularity, stability, and technical activity). Both studies offer complementary insights into the complex dynamics within FOSS.

It is evidenced from the surveyed studies that when FOSS communities' sustainability declines, several parts of their product quality or the mechanisms that assure it deteriorate as well. However, these studies either looked at one component of sustainability and its influence on quality in isolation or restricted the concept of quality to a certain metric (e.g., defect removal rate or new defect rate). Our research, on the other hand, investigates sustainability holistically (a collection of sustainability indicators as reported by Lin{\aa}ker et al. \cite{linaaker2022characterize}) and its influence on quality.

\section{Measuring FOSS sustainability}\label{sec:sustainability}

We adopted the framework developed by Lin{\aa}ker et al. \cite{linaaker2022characterize} for assessing the sustainability of FOSS. The framework presents an overview of a broad range of sustainability features. The framework presents a set of themes, and then within those themes, measurements are supplied under numerous emphasis areas, each of which has a distinct purpose that the metrics try to address. For example, under the theme ``communication,'' metrics like ``response quality,'' which evaluates the community's responsiveness (level of detail, complexity, and correctness), are shown by the community's communication channels like mailing lists, issue trackers, and pull requests \cite{linaaker2022characterize}. The framework is based on an analysis of 146 related publications.

The framework suggests 107 sustainability parameters organized into 15 themes \cite{linaaker2022characterize}. We rationalized the framework by reducing the themes and their associated parameters based on their prevalence in the literature. Within the ``communication'' theme, for example, ''response time'' is referenced more than ``visibility'' (the community's social media presence); whereas the latter is cited four times, the former is reported ten times \cite{linaaker2022characterize}. In addition, we de-scoped some themes and/or their corresponding parameters because of data unavailability; they cannot be sourced using our selected methods (see Sect. \ref{sec:methods}). For example, we de-scoped the theme ``culture,'' because its associated parameters, such as ``conflicts'' (the presence of conflicts in the community) \cite{linaaker2022characterize}, can only be sourced using qualitative methods. Although this rationalization may have discriminated indirectly against some sustainability parameters (see further discussion in Sect. \ref{sec:validity}, threats to validity), our reduced framework (Tbl. \ref{tbl:framework}) still remains extensive, with four themes and 16 parameters to assess FOSS sustainability.

We strategically opted for Lin{\aa}ker et al. \cite{linaaker2022characterize} to allow for a comprehensive and nuanced assessment of FOSS sustainability. The framework is distilled from an extensive literature on FOSS sustainability, offering a wide range of perspectives and indicators. By encompassing themes such as communication, popularity, stability, and technical activity, the framework allows for a holistic assessment of FOSS sustainability, which mirrors the complex and multi-layered nature of the phenomenon. The framework also allowed us to adequately and better capture FOSS sustainability, as opposed to simplistic or singular measures used in previous studies, e.g., \cite{foucault2015impact,wang2020unveiling}. Methodologically, our approach allowed us to cover multiple dimensions in our analysis, thereby contributing to a richer and more accurate representation of FOSS sustainability and how it influences software quality.

\section{Measuring software quality}\label{sec:swq}

\begin{table*}[t!]
    
    \footnotesize
    \vspace{2mm}
    
    \renewcommand\arraystretch{1.0}
    \caption{FOSS sustainability parameters adopted from Lin{\aa}ker et al. \cite{linaaker2022characterize}. ``Dev'' refers to development.}
    \label{tbl:framework}
  
    \begin{tabular}{p{2.2cm}lp{3.8cm}p{7.5cm}}
    
    \toprule
      \textbf{Theme} & \textbf{ID} & \textbf{Sustainability parameter} & \textbf{Definition}\\ \hline
      
        \multirow{4}{2.5cm}{Communication} & \multirow{2}{1cm}{COM-1} & \multirow{2}{3.1cm}{Response time} & The time elapsed between when a community member posted a comment/question and when he or she got a reply/response \cite{linaaker2022characterize}. \\ \cline{2-4}

        & \multirow{2}{1cm}{COM-2} & \multirow{2}{3.1cm}{Frequency of communication} & Frequency in FOSS project communication, such as the number of problems opened or comments posted in a certain time frame \cite{linaaker2022characterize}. \\ \hline
        
        \multirow{1}{2.5cm}{Popularity} & POP-1 & Project popularity & The projects' overall popularity, as reflected in forks, stars, and watchers \cite{linaaker2022characterize}.\\ \hline

        \multirow{14}{2.5cm}{Stability} & \multirow{1}{1cm}{STA-1} & \multirow{1}{3.1cm}{Age} & The age of the community is the duration in years from inception to current date \cite{linaaker2022characterize}.\\ \cline{2-4}
        
        & \multirow{2}{1cm}{STA-2} & \multirow{2}{3.1cm}{Attrition} & Attrition is the gradual reduction in contributors. This means that contributors are leaving faster than they are newcomers to the community \cite{linaaker2022characterize}.\\ \cline{2-4}

        & STA-3 & Forks & The number of project's forks. \cite{linaaker2022characterize}.\\ \cline{2-4}

        & \multirow{1}{1cm}{STA-4} & \multirow{1}{3.1cm}{Growth} & FOSS project growth and progress, as well as overall technological activity. \cite{linaaker2022characterize}.\\ \cline{2-4}

        & \multirow{2}{1cm}{STA-5} & \multirow{2}{3.1cm}{Knowledge concentration} & Distribution of contributions and expertise to specific persons or groups within an FOSS project, usually measured and explained using a community's bus- or truck factor \cite{linaaker2022characterize}. \\ \cline{2-4}

        & \multirow{1}{1cm}{STA-6} & \multirow{1}{3.1cm}{Life-cycle stage}  & The stage in the life cycle a FOSS project currently resides in. For example, growth, or dormancy \cite{linaaker2022characterize}.\\ \cline{2-4}

        & \multirow{2}{1cm}{STA-7} & \multirow{2}{3.1cm}{Retention} & The capacity of a FOSS community to keep its contributors active in the community for an extended length of time \cite{linaaker2022characterize}. \\ \cline{2-4}

        & \multirow{1}{1cm}{STA-8} & \multirow{1}{3.1cm}{Size} & The size of the FOSS project in terms of users and developers in any specific point in time \cite{linaaker2022characterize}.\\ \cline{2-4}

        & \multirow{2}{1cm}{STA-9} & \multirow{2}{3.1cm}{Turnover} & The number of contributors and maintainers who depart a FOSS community in a given time frame \cite{linaaker2022characterize}.\\ \hline

        \multirow{8}{2.5cm}{Technical activity} & \multirow{2}{1cm}{TEC-1} & \multirow{2}{3.1cm}{Contributors' dev. activity} & This value relates to the degree of non-maintainer contributions, such as development and technical writing. Looking at the entire effort less the labor of the maintainers \cite{linaaker2022characterize}.\\ \cline{2-4}

        & \multirow{2}{1cm}{TEC-2} & \multirow{2}{3.1cm}{Efficiency} & The effectiveness and convenience with which a FOSS project manages and advances development, such as accepting and evaluating issues and pull-requests \cite{linaaker2022characterize}. \\ \cline{2-4}

        & \multirow{2}{1cm}{TEC-3} & \multirow{2}{3.1cm}{Non-code contributions} & Technical activity of specifically non-code-contributions, i.e. documentation, community management, answering questions etc. \cite{linaaker2022characterize}.\\ \cline{2-4}

        & \multirow{2}{1cm}{TEC-4} & \multirow{2}{3.1cm}{Overall dev. activity} & Technical activity overall by the community, for example, development and technical writing. I.e. including both the maintainers' and contributors' work \cite{linaaker2022characterize}. \\ 
    
    \bottomrule
      
  \end{tabular}
  \vspace{-0.5cm}

\end{table*}

\noindent Software quality is the outcome variable we seek to understand in the context of FOSS sustainability. Software quality is a complex concept ``because it means different things to different people'' \cite{kitchenham1996software}. For example, ISO/IEC 25010 defines software quality as ``the degree to which the system satisfies the stated and implied needs of its various stakeholders and thus provides value'' \cite{iso2011}. The ISO/IEC 25010 model also proposes additional non-functional characteristics to include in the definition of software quality. These include ``functional suitability'', ``performance efficiency'', ``compatibility'', ``usability'', ``reliability'', ``security'', ``maintainability'', and ``portability'' \cite{iso2011}. Alami and Krancher interviewed practitioners on their expectations of the definition of software quality \cite{alami2022scrum}. They found that practitioners distinguish software quality characteristics into two discrete categories: ``external,'' which users can observe and evaluate, and ``internal,'' which developers assess during the software development and testing process \cite{alami2022scrum}. For example, ``conformity to business needs'' and software ``free of defects'' are external qualities of quality, and ``code maintainability'' and ``sustainable design'', which enable growth, are internal quality features \cite{alami2022scrum}. Some ISO/IEC 25010 standards' characteristics can straddle both external and internal categories. However, attributes such as ''security'' and ``usability'' can be considered external, and ``maintainability'' aligns well with internal. This parallel highlights the convergence between established standards and practice.

However, in the context of our study, we are constrained by Mining Software Repositories (MSR) techniques. Meeting the expectations of the ISO/IEC 25010 definition and those elucidated by practitioners in Alami and Krancher's study \cite{alami2022scrum,iso2011} may prove unfeasible. Using MSR, we relied on data that is readily available from software repositories. This inherent methodological constraint limits our ability to directly measure certain software quality attributes, such as ``usability'' and ``performance efficiency.'' These attributes may require runtime analysis for accurate assessment, which is beyond the scope of data accessible using MSR. Therefore, we focused on characteristics that can be mined and quantitatively analyzed using available repository data, e.g., defect density and code maintainability.

Similar studies in software engineering (SE) had different approaches to measuring software quality. While some, e.g., \cite{khomh2012faster,ray2014large,wang2020unveiling}, had a precise but narrow definition, i.e., \textit{``the number of bugs found during a project-month''} \cite{wang2020unveiling}, others, e.g., \cite{stuanciulescu2022code}, used more elaborate metrics to assess ``code quality,'' such as code coverage, functions that have a medium risk complexity (McCabe index between 11-25), number of functions that have a very high risk complexity (McCabe index $>$50), number of very large files ($>$1000 SLOC) \cite{stuanciulescu2022code}. In this study, we maintained the above-mentioned tradition of including metrics reported in SE literature. We used two metrics: defect density and code quality. Table \ref{tbl:swquality} reports the definitions of these metrics, and the computation methods are available in Sect.\,\ref{sec:methods}.

Our work exceeds the scope of similar studies assessing software quality. While similar studies relied on limited metrics, such as bug counts per project per month, our approach extends this by incorporating a broader array of quality attributes. We include defect density and also several code quality metrics, such as code coverage and various complexity metrics (e.g., McCabe index), see Tbl. \ref{tbl:swquality}. This approach allows us to include both internal attributes, influenced by developers, and external attributes, reflecting end users expectations of quality. By mining the project's issue trackers to identify defects, we incorporate defects reported by both developers and end-users, thereby including end-user perspectives. This inclusion addresses external quality attributes, as defects reported in issue trackers cover a broad array of defect types, including usability and performance issues. Our approach to some extent resonates with the ISO/IEC 25010 definition \cite{iso2011} and the insights from industry practitioners reported in Alami and Krancher's study \cite{alami2022scrum}.

\begin{table*}[th!]
    
    \footnotesize
    \vspace{2mm}
    
    \renewcommand\arraystretch{1.0}
    \caption{Software Quality Metrics Adopted in this Study}
    \label{tbl:swquality}
  
    \begin{tabular}{p{1.5cm}p{2cm}p{11cm}}
    
    \toprule
      \textbf{ID} & \textbf{Metric} & \textbf{Definition}\\ \hline
      
        \multirow{2}{1.5cm}{\textbf{SWQ-1}} & \multirow{2}{2cm}{Defect density} & The defect density is defined as the number of confirmed bugs in a program or module during a given time period of operation or development divided by the total size of the software or module (KLOC, size of the release, or per development day) \cite{shah2013software}.\\ \hline
        
        \multirow{4}{1.5cm}{\textbf{SWQ-2}} & \multirow{4}{2cm}{Code quality} & Code quality attributes ``not visible to the end-users'', i.e., ``internal'' and ``structural'' qualities \cite{alami2022scrum}. This metric encompasses code coverage ((Number of lines or statements tested / Total number of lines or statements) x 100\%) (\textbf{SWQ-2.1}), medium risk complexity (McCabe index between 11-25) (\textbf{SWQ-2.2}), very high risk complexity (McCabe index $>$50) (\textbf{SWQ-2.3}), number of very large files ($>$1000 SLOC) (\textbf{SWQ-2.4}), Very Large Function Size Count (\textbf{SWQ-2.5}), Code Duplication Percentage (\textbf{SWQ-2.6}), and Most Complex Function LOC (\textbf{SWQ-2.7}) \cite{stuanciulescu2022code}.\\ 
    
    \bottomrule
      
  \end{tabular}
  \vspace{-0.4cm}

\end{table*}
\section{Hypotheses Development}\label{sec:hypotheses}

In this study, inspired by the Lin{\aa}ker et al. \cite{linaaker2022characterize} framework, we propose that FOSS sustainability can be measured using four themes and their associated parameters: communication, popularity, stability, technical activity, and documentation \cite{linaaker2022characterize}. Drawing on this design, we develop hypotheses to test how these themes' parameters impact SWQ. In brief, we hypothesize that a decline in FOSS sustainability leads to a decrease in SWQ.

Although ``decline'' may signify a dynamic state suggesting FOSS sustainability is subject to fluctuations and changes over time, our computed metrics translate this dynamism into an accumulative assessment of community sustainability, whether it's in a state of decline or sustainability. This approach allowed us to quantitatively capture and summarize the evolving state of sustainability within FOSS communities. For instance, consider a FOSS project that has experienced growth and decline over time. If the project had experienced growth in its early years, our data would mirror this sustained upward trend in sustainability indicators. Subsequently, if the project encountered difficulties (for example, attrition) after the growth period, Then, our data would capture the accumulative impact of these fluctuations, offering an assessment of sustainability over time. This historical perspective allows a better understanding of the impact of the overall trajectory of FOSS community sustainability.

\subsection{Communication}

The main premise of the literature on this aspect of FOSS sustainability is that delivering high-quality communication to community contributors is critical to its survival \cite{wang2012survival}. Communication quality is defined as the combination of quality (degree of information and accuracy) \cite{linaaker2022characterize}, response time or responsiveness \cite{jiang2019metrics}, and frequency of communicating \cite{shaikh2019selecting}.

The literature postulates that a decline in communication quality leads to dissatisfied contributors \cite{steinmacher2019overcoming,wang2012survival,guizani2021long} and, as a consequence, community sustainability \cite{linaaker2022characterize}. This might also signify a drop in software quality as a result of the de-prioritization of quality standards owing to a lack of contributors. Quality communication is critical to ensuring that contributors feel appreciated, engaged, and encouraged to continue contributing \cite{van2017health}. In addition, when communication quality deteriorates, the chance of misunderstandings, disputes, and delays increases, which may have a detrimental influence on software development and quality. Under the communication theme, we adopted two sustainability parameters from Lin{\aa}ker et al.'s work: \emph{response time} (\textbf{COM-1}), and \emph{frequency of communication} (\textbf{COM-2}). We anticipate:

\medskip

\noindent\textit{\textbf{H1a:} A decline in FOSS communities response time is associated with a decrease in SWQ.}

\noindent\textit{\textbf{H1b:} A decline in FOSS communities freq. of communication is associated with a decrease in SWQ.}

\subsection{Popularity}

This theme relates to the overall external interest in the FOSS project or its overarching ecosystem \cite{linaaker2022characterize}. These characteristics aid in determining the level of popularity and adoption of a FOSS project among both existing and potential end-users and contributors \cite{linaaker2022characterize}. Zhou and Mockus argue that the growth of the contributor base is crucial for the community's sustainability \cite{zhou2012make}. They indicate that the likelihood of a developer transitioning into a long-term contributor role is correlated with their personal inclination and the surrounding environment. A positive community outlook and heightened project popularity serve to diminish the likelihood of leaving the community \cite{zhou2012make}. According to Osman et al., \cite{osman2021health} sustainable projects are those that have reached a high level of maturity and activity, boasting a sizable and engaged community of contributors, a high popularity, and low code complexity. They used the number of watchers to measure popularity. They conclude that communities with more popularity are more likely to be actively developed and maintained. Existing literature supports this view; for example, Jansen et al. \cite{jansen2014measuring} argue that a project's download count is an indicator of its sustainability. Borges et al. report that three out of four users consider the project's star rating, although they add that such metrics should be used cautiously \cite{borges2016understanding}. They note that one should investigate the evolution of stars over time and potential underlying reasons, for example, a fast growth in the number of stars is often a result of promotion on social sites and may have less to do with an active and sustainable development of the FOSS project. Acknowledging this limitation, testing it as a sustainability metric is a reliability test of this metric in itself. We hypothesize:

\medskip

\noindent\textit{\textbf{H2:} A decline in FOSS communities popularity is associated with a decrease in SWQ.}

\subsection{Stability}

The stability of the FOSS community is determined by its resilience and robustness, as well as the community's ability to maintain a consistent population of contributors who possess the necessary skills to sustain the FOSS project over an extended period of time \cite{linaaker2022characterize}. The literature extensively examines the growth, retention, attrition, and overall turnover of developers, as well as the concentration of contributions and knowledge among specific individuals or groups within the project, which is commonly measured and characterized by the bus or truck factor of the community as well as the size of the community \cite{linaaker2022characterize}. The potential decrease in contributors and resources available to support a project could result from a decline in the stability of a FOSS community. This may lead to a decelerating development process, reduced testing, and a decreased frequency of updates and bug fixes. Prior research has focused on analyzing stability indicators, such as turnover \cite{foucault2015impact} and elite developers \cite{wang2020unveiling}, in a singular manner. However, our study expands upon this approach by incorporating nine parameters, as suggested by Lin{\aa}ker et al. \cite{linaaker2022characterize}, to provide a more comprehensive analysis. Under the stability theme, we selected nine sustainability parameters: \emph{age} (\textbf{SAT-1}), \emph{attrition} (\textbf{SAT-2}), \emph{forks} (\textbf{SAT-3}), \emph{growth} (\textbf{SAT-4}), \emph{knowledge concentration} (\textbf{SAT-5}), \emph{life-cycle stage} (\textbf{SAT-6}), \emph{retention} (\textbf{SAT-7}), \emph{size} (\textbf{SAT-8}), and \emph{turnover} (\textbf{SAT9}). We expect:

\bigskip

\noindent\textit{\textbf{H3a:} The decrease in FOSS communities' age is associated with a decrease in SWQ.}

\noindent\textit{\textbf{H3b:} High attrition in FOSS communities is associated with a decrease in SWQ.}

\noindent\textit{\textbf{H3c:} An increase in FOSS communities forks is associated with a decrease in SWQ.}

\noindent\textit{\textbf{H3d:} A decline in FOSS communities growth is associated with a decrease in SWQ.}

\noindent\textit{\textbf{H3e:} High knowledge concentration in FOSS communities is associated with a decrease in SWQ.}

\noindent\textit{\textbf{H3f:} The dormancy of FOSS communities is associated with a decrease in SWQ.}

\noindent\textit{\textbf{H3g:} A decrease in FOSS communities retention is associated with a decrease in SWQ.}

\noindent\textit{\textbf{H3h:} A decrease in FOSS communities' sizes is associated with a decrease in SWQ.}

\noindent\textit{\textbf{H3i:} High turnover in FOSS communities is associated with a decrease in SWQ.}

\subsection{Technical activity}

The technical activity pertains to the attributes that delineate the comprehensive technical productions of a FOSS project in terms of advancing and enhancing its technical and non-technical outputs \cite{linaaker2022characterize}. Lin{\aa}ker and colleagues have identified five parameters related to technical activity: contributors' development activity, efficiency, maintainers' development activity, non-code contributions, and overall development activity \cite{linaaker2022characterize} (see Tbl. \ref{tbl:framework}). The aforementioned metrics highlight the significance of upholding a consistent and effective generation of software artifacts to ensure the sustainability of the community.

According to Midha and Palvia \cite{midha2012factors}, community ``technical success'' is based on the level of developer activity, which refers to the amount of effort put in by the contributors. The study revealed that a greater number of developers is positively correlated with increased developer activity \cite{midha2012factors}. The heightened level of activity serves as a promotional mechanism, drawing in new users and developers \cite{midha2012factors}. Sustaining a high level of activity can be achieved through the continuous addition of features and the resolution of bugs. Mockus et al. \cite{mockus2002two} posited that a substantial developer group is necessary for addressing bugs in FOSS projects beyond the core team. These findings have been corroborated by Champion and Hill \cite{champion2021underproduction}. The authors elucidate the concept of ``underproduction,'' which arises when the supply of labor becomes incongruous with the demand from users who depend on the software \cite{champion2021underproduction}. Their analysis reveals a prevalent issue of underproduction in various Debian software components that are commonly adopted.

It is contended that in instances where technical activity is minimal, it can prove to be arduous to sustain the requisite impetus to propel the project forward. In the absence of novel functionalities and the rectification of bugs, the project may experience stagnation. Insufficient technical engagement may result in the postponement of feature releases and bug fixes. Under this theme, we adopted four parameters: \emph{contributors' development activity} (\textbf{TEC-1}), \emph{efficiency} (\textbf{TEC-2}), \emph{non-code contributions} (\textbf{TEC-3}), and \emph{overall development activity} (\textbf{TEC-4}). We anticipate:

\medskip

\noindent\textit{\textbf{H4a:} A decline in FOSS communities contributors' dev. activity is associated with a decrease in SWQ.}

\noindent\textit{\textbf{H4b:} A decline in FOSS communities efficiency is associated with a decrease in SWQ.}

\noindent\textit{\textbf{H4c:} A decline in FOSS communities non-code contributions is associated with a decrease in SWQ.}

\noindent\textit{\textbf{H4d:} A decline in FOSS communities overall dev. activity is associated with a decrease in SWQ.}

\smallskip

The examination of these hypotheses will assist us in understanding how these sustainability themes (communication, popularity, stability, technical activity, and documentation) impact an important community outcome, i.e., quality \cite{vasilescu2015quality}. This will provide some insights on how far-reaching the influence of FOSS sustainability is in changing the capacity to uphold quality standards.
\section{Methods}\label{sec:methods}

\subsection{Data collection \& computation}

We selected the ASFI projects to carry out the study. This choice is motivated by the unique opportunity the AFSI data offer to understand the complex phenomenon of sustainability and its impact on software quality. The ASF incubator hosts a diverse range of projects spanning various domains, from infrastructure and data processing to web development and artificial intelligence. This diversity provides an opportunity to study sustainability and quality factors across different types of software projects, making the findings more applicable to a broader spectrum of FOSS initiatives. The ASF Incubator also hosts projects at various stages of development and maturity. This enables us to analyze projects from their early stages to more mature and sustainable states, providing more credibility and robustness to our claims.

We selected the projects using St{\u{a}}nciulescu, et al.'s \cite{stuanciulescu2022code} list. The list is still recent and relevant. Since the St{\u{a}}nciulescu, et al. list is well documented, it facilitates data retrieval and ensures transparency in project selection. This transparency is essential for reproducibility and for building upon their findings in our study. However, not all listed projects fully met our selection requirements due to project setup and infrastructure constraints. Initially, we started with 236 projects from the original list. We then excluded projects without PRs in GitHub or Jira, empty issue trackers, non-GitHub/Jira issue hosting, and unavailable defect labels (essential for \textbf{SWQ-1} computation). Our final list includes 217 projects, down from the original 236 in St{\u{a}}nciulescu, et al.'s list.

We employed specific tools and developed Python scripts to ensure the systematic retrieval of projects' data. The data download occurred iteratively in the 1st quarter of 2023, mainly attributed to download restrictions enforced by the GitHub and Jira platforms. We utilized the git tool to clone all projects and the \href{https://github.com/chaoss/grimoirelab-perceval}{Perceval}\footnote{\url{https://github.com/chaoss/grimoirelab-perceval}} tool to aggregate data components like commits, issues, and PRs, along with repository-related information (for example, stars, watchers, forks, and repository size in KB). Our Python scripts served to execute the following actions: clone repositories from GitHub; extract commits from the cloned repositories; download issues, PRs, and repository information from GitHub; and acquire issues from Jira. To maintain a cohesive data structure across all projects, we standardized the storage approach. Collectively, our efforts amassed a dataset exceeding 100 GB (35 GB in a compressed tarball format). After running Sokrates to compute SWQ metrics on the downloaded git repositories, the size of the collected data increased to 159 GB of uncompressed and 39 GB of compressed tarball.

For the computation of our metrics, we balanced between the adherence to the original definition in the literature, as presented by Lin{\aa}ker and colleagues \cite{linaaker2022characterize}, and the constraints inherent to the repository data. However, we contend that our computation has preserved the authenticity of the original definitions to a great extent (see section \ref{sec:validity} for further discussion). We initiated our metrics computation from the inception of the repository (start date) and concluded it at the most recent date of project activity (end date), such as an issue or a PR created.

For \textbf{COM-1} (\emph{response time}), we calculated the average time it takes for the first comment to appear for an issue. \textbf{COM-2} (\emph{frequency of communication}) is the number of comments in all issues plus the total number of issues. While PR comments are valuable for tracking code-related discussions and reviewing contributions, they represent a narrower subset of community interactions. We emphasized issue comments over PR comments because issues serve as the entry point for identifying problems, enhancements, and features, making them the heart of community engagement, which fits with the original intent of these metrics (see Tbl. \ref{tbl:framework}). Issue comments, therefore, provide a direct window into how promptly and effectively issues are acknowledged, discussed, and resolved. This reflects the community's responsiveness to user concerns and its ability to efficiently manage and prioritize tasks. PR comments may not present the broader context of the project's health, including the handling of non-code issues, user feedback, and strategic decision-making. \textbf{POP-1} (\emph{project popularity}) is the combined total of forks, stars, and watchers.

\textbf{STA-1} (\emph{age}) is the age in years, and it is calculated from the date of the download since the project inception. For \textbf{STA-2} (\emph{attrition}), we used the number of commits over the lifespan of projects; i.e., if the number of commits decreases, then there is a reduction in contributions. So, \textbf{STA-2} is the cumulative decrease in the number of commits, in periods of twelve weeks, during a specified time span (from the project's inception date until the most recent date of project activity, such as a PR created). \textbf{STA-3} (\emph{forks}) is the number of forks. \textbf{STA-4} (\emph{growth}) is to determine the project's growth and progress, as well as overall technological activity. While this definition cannot translate directly to a concise formula, as ``growth and progress'' could be evaluated in various ways, nevertheless, we used the growth in PR submissions to compute this metric. So, \textbf{STA-4} represents the cumulative increase in the number of PRs, in periods of twelve weeks, during a specified time span (from the project's inception date until the most recent date of project activity, such as an issue created). We adopted Avelino et al.'s definition of the ``truck factor,'' which implies the minimum number of contributors who need to depart a project before it becomes compromised \cite{avelino2016novel}. To compute the \textbf{STA-5} (\emph{knowledge concentration}), metric, we use a tool\footnote{\url{https://github.com/aserg-ufmg/Truck-Factor}} made available by the same research team \cite{avelino2016novel}. We adopted Valiev et al.'s definition of dormancy, ``having very little or no development activity after some time'' \cite{valiev2018ecosystem}, to compute \textbf{STA-6} (\emph{life-cycle stage}). To this end, we compute the average number of commits per month in the last twelve months from our end date (the most recent date of project activity). If the number is below zero, then the project is considered dormant.

We computed \textbf{STA-7} (\emph{retention}) by calculating the cumulative total of annual increases in the number of active contributors. We used Wang et al.'s \cite{wang2020unveiling} method to determine ``active'' contributors. The study sets a time period of 90 days as the duration of the ``elite-ship'' \cite{wang2020unveiling}. If a developer performs a task within the 90-day period that requires write permission (i.e., code changes), their ``elite-ship'' status is renewed for another 3 months, starting from the month in which they performed the task. This renewal indicates that they remain active and capable of making code changes. To implement this, we segmented our designated time period into consecutive 3-month snapshots, enabling us to closely observe contributors' actions, specifically their commits, within each snapshot. If a contributor engages in the repository by making at least one commit during a particular 3-month period, they are considered to be maintaining their active status within that timeframe. \textbf{STA-8} (\emph{size}) represents the project's size in terms of its user and developer community at any given point in time \cite{linaaker2022characterize}. We calculated STA-8 by counting contributors who have engaged in at least one commit, PR, issue, or issue comment. It is important to note that each contributor was counted only once, even if they were involved in multiple activities within the project. However, due to the way we tracked contributions, there are instances where contributors might appear more than once in the count. This can occur when the same author uses different identifiers, such as email addresses or GitHub handles, for issues and PRs. Additionally, if users change their identifiers during the specified time interval, they could also be counted more than once (corresponding to each identifier change). \textbf{STA-9} (\emph{turnover}) represents the count of contributors who authored commits and have been inactive in the preceding six-month period, as measured from the designated end date.

\textbf{TEC-1} (\emph{contributors' development activity}) is the total count of commits made by non-maintainers. We define a non-maintainer as a contributor who has not yet merged any PR. \textbf{TEC-2} (\emph{efficiency}) measures the time elapsed from PR creation until it is merged or closed. \textbf{TEC-3} (\emph{non-code contributions}) is the count of commits of files that are not related to programming code; the file formats we considered are ``txt'' and ``md.'' On the other hand, \textbf{TEC-4} (\emph{overall development activity}) represents coding contributions, the count of commits of coding files (other than ``txt'' and ``md'').

The \textbf{SWQ-1} (\emph{defect density}) is the ratio of defects, labeled as ``defects'' in the issues list, to the size of the project measured in kilobytes. We manually inspected the issues for each project and extracted their labels. Then, we use those labels to identify the defect issues. We followed the Stuanciulescu et al. \cite{stuanciulescu2022code} method to determine \textbf{SWQ-2} (\emph{code quality}). We used \href{https://www.sokrates.dev/}{Sokrates}\footnote{\url{https://www.sokrates.dev/}} to extract various sub-metrics (see Tbl. \ref{tbl:swquality}) related to \textbf{SWQ-2} for each incubation month. We tested \textbf{SWQ-2} sub-metrics separately: \textbf{SWQ-2.1} (\emph{code coverage}), \textbf{SWQ-2.2} (\emph{medium risk complexity}), \textbf{SWQ-2.3} (\emph{very risk complexity}), \textbf{SWQ-2.4} (\emph{number of very large files}), \textbf{SWQ-2.5} (very large function size count), \textbf{SWQ-2.6} (code duplication percentage), and \textbf{SWQ-2.7} (most complex function).

%%% macros %%%
\newcommand{\Issues}{\ensuremath{I}}
\newcommand{\Comments}{\ensuremath{C}}
\newcommand{\Forks}{\ensuremath{F}}
\newcommand{\Stars}{\ensuremath{S}}
\newcommand{\Watchers}{\ensuremath{W}}
\renewcommand{\time}{\ensuremath{\mathit{time}}} % already defined by some package, but I suppose it does not matter to re-define it...
\renewcommand{\year}{\ensuremath{\mathit{year}}} % already defined by some package, but I suppose it does not matter to re-define it...
\newcommand{\repo}{\ensuremath{\mathit{r}}}
\newcommand{\Commits}{\ensuremath{CM}}
\newcommand{\PullRequests}{\ensuremath{\mathit{PR}}}
\newcommand{\period}{\ensuremath{\mathit{period}}}
\newcommand{\periodinit}{\ensuremath{\mathit{init}}}
\newcommand{\periodend}{\ensuremath{\mathit{end}}}
\newcommand{\periodfreq}{\ensuremath{\mathit{freq}}}
\newcommand{\contributors}{\ensuremath{\mathit{contributors}}}
\newcommand{\activecontributors}{\ensuremath{\contributors_{\mathit{act}}}}
\newcommand{\contributor}{\ensuremath{\mathit{contr}}}
\newcommand{\commit}{\ensuremath{\mathit{cmt}}}
\newcommand{\mergers}{\ensuremath{\mathit{mergers}}}
\newcommand{\mergedclosed}{\ensuremath{\time_{\mathit{merged/closed}}}}
\newcommand{\pullrequest}{\ensuremath{\mathit{pr}}}
\newcommand{\files}{\ensuremath{\mathit{files}}}
\newcommand{\file}{\ensuremath{\mathit{f}}}
\newcommand{\Docs}{\ensuremath{\mathit{Docs}}}

%%%%% In table 3
\newcommand{\Decrements}{\ensuremath{\mathit{Decs}}}
\newcommand{\Increments}{\ensuremath{\mathit{Incs}}}
\newcommand{\NumberCommitsPeriod}{\ensuremath{\mathit{NC}}}
%%% macros %%%

Table \ref{tbl:smetrics} outlines the final formulas we use to compute our selected sustainability parameters. In what follows, we describe the notation employed in the table. We use $\Issues$ to represent the set of issues within a project. $\Comments_i$ denotes the set of comments in an issue $i \in \Issues$, with $c_i \in \Comments_i$ specifying a concrete comment under that issue; $C \triangleq \cup_{i \in I} C_i$ encompasses all comments across project issues. Chronological ordering between issues and comments is denoted with superscripts, for example, $i^1$ for the initial issue and $c^3_i$ for the third comment on issue $i$. The $\time$ function retrieves creation time for issues and comments. The size of a set $S$ is denoted as $|S|$. The quantities $|\Forks|$, $|\Stars|$, and $|\Watchers|$ indicate the number of forks, stars, and watchers, respectively. We use $\repo$ to refer to a project's repository, and $\year$ provides its inception year. $\Commits^t$ denotes a set of commits within a time period $t$; for example, $\Commits^{2023\text{-}05}$ denotes commits in May 2023, while $\Commits$ spans all project commits. The function $\period(\periodinit, \periodend, \periodfreq)$ yields a set of time periods within the start and end dates and a frequency (in weeks); for example, $\period(\text{2020-01-01},\text{2021-01-01},4)$ gives all 4-week periods from 2020-01-01 to 2021-01-01. Given a time period $t$, we use $t_\periodinit$ and $t_\periodend$ to refer to the beginning and end dates of the time period. $\PullRequests^t$ refers to the set of PRs in time period $t$. We use \mergedclosed\ to retrieve the time when a PR was merged or closed. The function $\contributors$ takes as input a set of either commits, pull requests, issues, or comments and returns the set of contributors who created the commits, pull requests, issues, or comments, respectively. Similarly, \activecontributors\ returns the set of active contributors in a given a set of commits. Furthermore, given a contributor $\contributor$ and a time period $t$, we use $\Commits^t_\contributor$ to denote the commits of \contributor\ during time period $t$. The function $\mergers$ takes as input a set of pull requests and returns the set of contributors who merged them. Given a commit $\commit$, the function $\files(\commit)$ returns all the files modified in $\commit$. Finally, $\Docs$ denotes the set of documentation files in a project.

\begin{table*}[t]
    
    \footnotesize
    \vspace{2mm}
    
    \renewcommand\arraystretch{1.0}
    \setlength\extrarowheight{3mm}
    \caption{FOSS sustainability parameters computation methods}
    \label{tbl:smetrics}
  
    \begin{tabular}{m{1.5cm}m{4.0cm}m{9.5cm}@{}}
    
      \toprule
      \textbf{ID} 
      & \textbf{Sustainability parameter} 
      & \textbf{Computation method}
      \\ \midrule
      
      \textbf{COM-1} 
      & Response time 
      & $\sum_{i \in I} (\time(c^1_i) - \time(i)) / |I| $ 
      \\ %\hline

      \textbf{COM-2} 
      & Frequency of communication 
      & $|\Comments| + |\Issues|$
      \\ %\hline
        
      \textbf{POP-1}
      & Project popularity
      & $|\Forks| + |\Stars| + |\Watchers|$
      \\ %\hline

      \textbf{STA-1}
      & Age
      & $2023 - \year(\repo)$
      \\ %\hline
        
      \textbf{STA-2}
      & Attrition
      & $ \sum_{d \in D} d \text{ with } D = \{ |\Commits^t| - |\Commits^{t+1}| \mid \Commits^j \in \Commits, t + 1, t \in \period(\periodinit,\periodend,\periodfreq) \}$
      \\ %\hline

      \textbf{STA-3}
      & Forks
      & $|\Forks|$
      \\ %\hline

      \textbf{STA-4}
      & Growth
      & $ \sum_{i \in \Increments} i \text{ with } \Increments \triangleq \{ |\PullRequests^{t+1}| - |\PullRequests^t|  \mid  \PullRequests^j \in \PullRequests. t + 1, t \in \period(\periodinit,\periodend,\periodfreq) \}$
      \\ % \hline

      \textbf{STA-5}
      & Knowledge concentration
      & Algorithm 1 in \cite{avelino2016novel}
      \\ % \hline

      \textbf{STA-6}
      & Life-cycle stage
      & $   \sum_{n \in \NumberCommitsPeriod} n / |\NumberCommitsPeriod| \text{ with } \NumberCommitsPeriod \triangleq \{ |\mathit{CM}^t \in \mathit{CM}|  \mid  t \in \period(\periodinit - \textit{1 year}, \periodend, 4) \} < 1$
      \\ % \hline

      \textbf{STA-7}
      & Retention
      & $   \sum_{i \in \Increments} i \text{ with } \Increments \triangleq \{ |\activecontributors(\Commits^{t+1})| - |\activecontributors(\Commits^t)|  \mid  \Commits^j \in \Commits, t + 1, t \in \period(\periodinit,\periodend,\periodfreq)  \}$
      % & $   \sum_{i \in \Increments} i \text{ with } \Increments \triangleq \{ |\PullRequests^{t+1}| - |\PullRequests^t|  \mid  \PullRequests^j \in \PullRequests, t + 1, t \in \period(\periodinit,\periodend,\periodfreq)  \}$
      \\ % \hline

      \textbf{STA-8}
      & Size
      & $|\contributors(\Commits) \cup \contributors(\PullRequests) \cup \contributors(I) \cup \contributors(C)|$
      \\ % \hline

      \textbf{STA-9}
      & Turnover
      & $|\{ \contributor \in \contributors(\Commits^t) \mid \exists \commit \in \Commits^t_{\contributor} \cdot t_\periodend - \time(\commit) < \textit{6 months} \}|$
      \\ % \hline

      \textbf{TEC-1}
      & Contributors' dev. activity
      & $|\{\contributor \in \contributors(\Commits) \mid \contributor \not \in \mergers(\PullRequests) \}|$
      \\ % \hline

      \textbf{TEC-2}
      & Efficiency
      & $\sum_{t \in T} t/|\PullRequests| \text{ with } T = \{\mergedclosed(\pullrequest) - \time(\pullrequest) \mid \pullrequest \in \PullRequests\}$
      \\ % \hline

      \textbf{TEC-3}
      & Non-code contributions
      & $|\{ \commit \in \Commits \mid \exists \file \in \files(\commit) \cdot \file \in \Docs \}|$
      \\ % \hline

      \textbf{TEC-4}
      & Overall development activity
      & $|\{ \commit \in \Commits \mid \forall \file \in \files(\commit) \cdot \file \not \in \Docs \}|$
      \\ 
    
    \bottomrule
      
  \end{tabular}

\end{table*}

\subsection{Data analysis}
\label{subsec:data-analysis}

We follow a Bayesian approach for data analysis. This provides a more nuanced and probabilistic approach compared to traditional frequentist methods \cite{furia2022applying}. Bayesian analysis begins with prior distributions, which summarize plausible parameter values before observing the data. Collected data are included as observations. Then, Bayesian inference is used to redistribute probability over parameter values according to the observations. The outcome is an update of the original prior beliefs, i.e., a posterior distribution over parameter values~\cite{kruschke2018bayesian}. The 95\% High Density Interval (HDI) of the posterior distribution comprises parameter values whose cumulative probability amounts to 95\% of probability density. These parameter values are considered the most credible and viable findings. Parameter values falling in the remaining 5\% cumulative probability are considered non-credible~\cite{kruschke2018bayesian}.

Overall, posterior distributions summarize the relative credibility of all possible parameter values. In our analysis, posterior distributions provide a detailed view of how strongly sustainability metrics impact quality metrics (or the lack thereof). Furthermore, we complement our analysis with a binary answer to our hypothesis. 

To this end, we define a Bayesian decision criterion based on the HDI of the posterior distribution of parameter values. To determine whether a sustainability metric has an impact on quality metrics, we should find statistical evidence that parameter values for this metric are distinct than 0. Hence, the decision criterion is as follows:

\begin{itemize}
    \item [-] If the HDI of a parameter includes 0, we conclude that there is no statistical evidence to assert an impact of the target sustainability metric on quality.

    \item [-] If the HDI does not include 0, we conclude that the sustainability metric has an impact on the target quality metric. In particular, an HDI whose values are greater than 0 indicates that an increase in the values of the sustainability metric entails an increase of the values of the quality metric. Analogously, if HDI  values are less than 0, then an increase in the values of the sustainability metric entails a decrease in the values of the quality metric.
    
    % \item [-] If the HDI does not include 0, we conclude that there is either a negative or positive impact the sustainability metric on quality. In particular, an HDI whose values are greater than 0 indicates statistical evidence for a positive impact. Analogously, an HDI with values less than 0 indicate a negative impact.
    
\end{itemize}
  
The validity of this criterion lies at the definition of HDI. When 0 is included in the HDI, it becomes a plausible value that cannot be discarded; which provides statistical evidence for the sustainability metric having no impact on quality. However, when 0 is outside the HDI, it is not considered a credible value---as its associated probability is less than 5\%. This, in turn, provides evidence for a positive/negative impact. Due to the units of each metric and data transformations, the strength of the impact is to be interpreted on a case-by-case basis.

We consider two standard types of Bayesian probabilistic models, depending on the scale of the target software quality variable. For \textbf{SWQ-1}, \textbf{SWQ-2.1}, and \textbf{SWQ-2.6}, since they are on a metric scale, we use Gaussian regression. The remaining software quality variables contain count data (natural numbers including 0). Thus, we use Poisson regression. These are standard models for these types of variables~\cite{kruschke2018bayesian}. The structure of the model types is as follows (left: Gaussian regression, right: Poisson regression):

\[
\begin{aligned}
\textit{\underline{Gaussian regression}} \hspace{-5cm}&\\
\sigma & \sim \mathcal{U}(10^{-3}, 10) \\
\alpha_i & \sim \mathcal{N}(0, 10) \quad \text{for } i \in \{\text{COM-1}, \ldots, \text{TEC-4}\} \\
\delta_j & \sim \mathcal{N}(0, 1) \quad \text{for } j \in \{\text{non-dormant}, \text{dormant}\} \\
\beta & \sim \mathcal{N}(0, 10) \\
\mu & = \beta + \delta_j + \sum_{i} \alpha_i x_i \\
\log(y) & \sim \mathcal{N}(\mu, \sigma) \\
\end{aligned}
\hspace{3mm}
\begin{aligned}
\textit{\underline{Poisson regression}} \hspace{-4.5cm}&\\
\alpha_i & \sim \mathcal{N}(0, 10) \quad \text{for } i \in \{\text{COM-1}, \ldots, \text{TEC-4}\} \\
\delta_j & \sim \mathcal{N}(0, 1) \quad \text{for } j \in \{\text{non-dormant}, \text{dormant}\} \\
\beta & \sim \mathcal{N}(0, 10) \\
\log(\lambda) & = \beta + \delta_j + \sum_{i} \alpha_i \log(x_i) \\
y & \sim \mathcal{P}(\lambda) \\
\end{aligned}
\]

Our analysis focuses on the impact of each individual sustainability metric on software quality. Therefore, these model types produce multiple model instances, one for each sustainability metric. We use $y$ to denote the outcome software quality data and $x_i$ the sustainability metric data.
We remark that this analysis differs from computing a correlation coefficient between variables. Correlation coefficients measure the extent to which two variables have a linear relation, e.g., Pearson coefficient (a popular regression coefficient) is a value in $(-1,1)$. Also, correlation coefficients are often symmetric meaning that they do not measure the impact of one variable on the other. They only measure their linear relation. Our Bayesian models are designed to measure the impact of a predicted variable (a sustainability metric) to a predicted variable (a software quality metric). The scale $\alpha_i$ parameter for predictors measures the strength of this impact, i.e., how much a change in sustainability increases quality. This is different from measuring the strength of the linear relation between the variables, which does not have such a direct interpretation. Thus, our Bayesian analysis provides an easy to interpret and very detailed analysis of the impact of sustainability metrics on quality compared to computing a correlation coefficient. In what follows, we describe the specific details of each model type.

\paragraph{Gaussian regression}
We start by performing two data transformations: i) log transform outcome variables and ii) data standardization for predictors. Data standardization consists in performing a linear transformation so that the data have a mean of zero and a standard deviation of one~\cite{kruschke2018bayesian}. This transformation is applied to $x_i$. This process improves the performance of Bayesian inference as the set of possible values for parameters is more concentrated. Computing the log of outcome quality metrics, $\log(y)$, helps to better fit the Gaussian distribution. As a result, we obtain better predictive models for Gaussian regression. The parameter $\alpha_i$ captures whether sustainability metrics have a positive or negative impact on the outcome variable, except for \textbf{STA-6}. Since \textbf{STA-6} is a binary categorical variable, it is modeled as two parameters, $\delta_j$, each of which determines the effect of \textit{non-dormant} and \textit{dormant} projects, respectively. The parameter $\beta$ is the intercept of the model, and $\sigma$ is the standard deviation of the Gaussian data distribution. 
We use a Gaussian, $\mathcal{N}(0,10)$, prior for $\alpha_i, \beta$.
These are non-informative priors that extensively cover all plausible values of the log-transformed outcome variable.
The prior on $\delta_j$ has lower standard deviation, $\mathcal{N}(0,1)$, to prevent undesired interactions with $\beta$ when estimating its value.
We use a uniform prior on positive values, $\mathcal{U}(10^{-3},10)$ for $\sigma$, as we must ensure $\sigma > 0$.
Due to the log transformation of the outcome, model parameters indicate the percentage increase or decrease in the quality metric per unit of the sustainability metric. Since this data is standardized, a unit increase corresponds to one standard deviation increase. Standard deviations for each metric are reported in the accompanying replication package.
 
To study the individual effect of each sustainability metric, we instantiate this model type with $\mu = \beta + \alpha_i x_i$ or $\mu = \delta_j$, depending on the sustainability metric under analysis. Since this model applies to 3 software quality metrics and all sustainability metrics, we study $3 \cdot 16 = 48$ models of this type.

\paragraph{Poisson regression}
Due to the intricacies of this type of model, we cannot perform the data standardization process as in Gaussian regression. Instead, we perform a log transformation of the data for sustainability metrics, $\log(x_i)$, and keep quality metrics, $y$, unmodified. The log transformation allows us to concentrate the range of possible values for the sustainability parameters $\alpha_i$. However, due to the large number of negative values and zeros in \textbf{STA-4}, \textbf{STA-7}, and \textbf{TEC-1}, we discard them for this type of model; recall log transforms are undefined for these values. The meaning of the parameters $\alpha_i$, $\delta_j$, and $\beta$ is the same as before. 
As priors, we use Gaussian distributions centered at 0, with a standard deviation of 10 (for $\alpha_i$ and $\beta$) and 1 (for $\delta_j$). As before, these are non-informative priors. Note that, as usual for Poisson regression, we use a logarithmic link function for the rate of the Poisson data distribution $\log(\lambda)$~\cite{kruschke2018bayesian}. Thus, our prior on the intercept $\beta$ covers a range $e^{-20} \approx 0$ to $e^{20} \approx 4.85 \times 10^8$. This range captures all plausible values for the target quality metrics and ensures that they are assigned non-zero prior probability. The log transformation of predictors implies that a $10\%$ increase in sustainability changes  $0.1\alpha_i$ or $0.1\delta_j$ units in the code quality metric.

As before, Poisson regression models are instantiated for single sustainability metrics as $\log(\lambda) = \beta + \delta_j$ for \textbf{STA-6} and $\log(\lambda) = \beta + \alpha_i x_i$ for the rest. This model applies to 5 software quality metrics and 13 sustainability metrics; hence, we study $5 \cdot 13 = 65$ models of this type.

To perform Bayesian inference, we implemented our models in the probabilistic programming library PyMC~\cite{pymc}. Since all the model parameters are continuous, we use the NUTS sampler, which is the best performing sampler for estimating continuous parameters~\cite{nuts}. For each model, we computed four parallel chains of 3000 samples each. All our analyses show a Monte Carlo Standard Error (MCSE) below 0.02, which indicates high accuracy~\cite{kruschke2018bayesian}. Furthermore, we also perform a posterior predictive check, which shows that our models have good predictive accuracy.

\section{Findings}\label{sec:findings}

Table \ref{tbl:result} summarizes the results; column \textbf{SWQ-1} lists the results for defect density and \textbf{SWQ-2.1 ... SWQ-2.7} for code quality metrics. The \ding{56} indicates that we found no impact of a particular metric on \textbf{SWQ-1} and/or \textbf{SWQ-2} sub-metrics (the metric in question results in no statistical evidence to confirm a positive or negative impact on SWQ metrics and sub-metrics), $+$ a positive impact (the increase of the metric in question was associated with better performance of SWQ metrics or sub-metrics), and $-$ a negative impact (the metric value increase was linked to a decrease in SWQ metrics or sub-metrics). For \textbf{STA-2} (\emph{attrition}), \textbf{STA-5} (\emph{knowledge concentration}), \textbf{STA-9} (\emph{turnover}), and \textbf{TEC-2} (\emph{efficiency}), a decrease in their values is associated with an improvement in SWQ metrics, denoted by a $+$ sign, indicating a positive impact. A $-$ sign is a negative impact; a decrease in their value is linked to a decrease in SWQ metrics. Recall that we discarded some sustainability metrics in Poisson regression models due to restrictions on the required data transformations (see Sect.~\ref{subsec:data-analysis}). When that was the case, we marked \textbf{NA} in the table.

Overall, our selected sustainability metrics show no evidence of a consistent positive or negative impact on our selected software quality metrics. However, even when exhibiting signs of sustainability, the communities in our sample sometimes demonstrated subpar results in some code quality metrics, such as very high risk complexity (\textbf{SWQ-2.3}), the number of very large files (\textbf{SWQ-2.4}), the very large function size count (\textbf{SWQ-2.5}), and the complexity of functions (\textbf{SWQ-2.7}). On the other hand, age (\textbf{STA-1}) has a positive impact on some code quality sub-metrics, except for \textbf{SWQ-2.2} (medium risk complexity). In the upcoming subsections, we discuss the results of our hypothesis tests, organized by themes: communication, popularity, stability, and technical activity.

For the sake of brevity and to ensure a concise presentation, we will use the \textbf{H1a - Response time} subsection to illustrate our detailed analytical approach conducted for all hypotheses. Then, for the remaining hypotheses, we will succinctly present the analysis, results, and our interpretations without delving into extensive procedural specifics. This choice has also allowed us to focus on the interpretations to facilitate a clearer, digestible, and more accessible synthesis of our research outcomes.

\begin{table*}[th!]
    
    % \small
    \vspace{2mm}
    \centering
    \renewcommand\arraystretch{0.65}
    \caption{Analysis results, \ding{56} (no impact), $\boldsymbol{+}$ (positive impact), $\boldsymbol{-}$ (negative impact), and \textbf{NA} (not computed). Due to log-transformations in the data, the initial sample of 217 projects fluctuated, with some projects being excluded due to values of 0 or negatives in certain metrics. The numbers adjacent to SWQ metrics represent the projects included in our analysis for all sustainability measures. If this number changed for a specific metric, the adjacent cell shows the final number.}
    % \vspace{-0.3cm}
    \label{tbl:result}
    {%\normalsize
    % \begin{tabular}{p{1.8cm}cccccccc}
    \hspace{-8mm}
    \begin{tabular}{
        p{3.1cm}
        m{0.9cm}
        m{0.9cm}
        m{0.9cm}
        m{0.9cm}
        m{0.9cm}
        m{0.9cm}
        m{0.9cm}
        m{0.9cm}
    }
    % \toprule
    
    %   \textbf{Metrics} & 
    %   & 
    %   \multicolumn{7}{c}{\textbf{SWQ-2}} \\       
    %   \cmidrule{3-9}
      % \textbf{Metrics} &
      &
      \rotatebox[origin=c]{30}{\textbf{SWQ-1} {\tiny (209)}}   & 
      \rotatebox[origin=c]{30}{\textbf{SWQ-2.1} {\tiny (211)}} & 
      \rotatebox[origin=c]{30}{\textbf{SWQ-2.2} {\tiny (215)}} & 
      \rotatebox[origin=c]{30}{\textbf{SWQ-2.3} {\tiny (215)}} & 
      \rotatebox[origin=c]{30}{\textbf{SWQ-2.4} {\tiny (215)}} & 
      \rotatebox[origin=c]{30}{\textbf{SWQ-2.5} {\tiny (215)}} & 
      \rotatebox[origin=c]{30}{\textbf{SWQ-2.6} {\tiny (213)}} & 
      \rotatebox[origin=c]{30}{\textbf{SWQ-2.7} {\tiny (215)}} \\ 
      \midrule
      
        \textbf{COM-1} (\textit{Response}) & \ding{56} & \ding{56} & \textbf{$\boldsymbol{+}$}  & \textbf{$\boldsymbol{-}$} & \textbf{$\boldsymbol{-}$} & \textbf{$\boldsymbol{-}$} & \ding{56} & \textbf{$\boldsymbol{-}$} \\ 

        \textbf{COM-2} (\textit{Freq. of com}) & \textbf{$\boldsymbol{-}$} & \textbf{$\boldsymbol{+}$} & \ding{56} & \textbf{$\boldsymbol{-}$} & \textbf{$\boldsymbol{-}$} & \textbf{$\boldsymbol{-}$} & \ding{56} & \textbf{$\boldsymbol{-}$} \\ \midrule
        
        \textbf{POP-1} (\textit{Popularity}) & \ding{56} & \textbf{$\boldsymbol{+}$} & \textbf{$\boldsymbol{+}$} & \textbf{$\boldsymbol{-}$} & \textbf{$\boldsymbol{-}$} & \textbf{$\boldsymbol{-}$} & \ding{56} & \textbf{$\boldsymbol{-}$}\\ \midrule

        \textbf{STA-1} (\textit{Age}) & \textbf{$\boldsymbol{-}$} & \ding{56} & \textbf{$\boldsymbol{-}$} {\tiny (170)} & \textbf{$\boldsymbol{+}$} {\tiny (170)} & \textbf{$\boldsymbol{+}$} {\tiny (170)} & \textbf{$\boldsymbol{+}$} {\tiny (170)} & \textbf{$\boldsymbol{+}$} & \textbf{$\boldsymbol{+}$} {\tiny (170)}\\ 

        \textbf{STA-2} (\textit{Attrition}) & \ding{56} & \ding{56} & \textbf{$\boldsymbol{+}$} & \textbf{$\boldsymbol{-}$} & \textbf{$\boldsymbol{-}$} & \textbf{$\boldsymbol{-}$} & \ding{56} & \textbf{$\boldsymbol{-}$}\\

        \textbf{STA-3} (\textit{Forks}) & \ding{56} & \textbf{$\boldsymbol{+}$} & \textbf{$\boldsymbol{+}$} & \textbf{$\boldsymbol{-}$} & \textbf{$\boldsymbol{-}$} & \textbf{$\boldsymbol{-}$} & \ding{56} & \textbf{$\boldsymbol{-}$}\\

        \textbf{STA-4} (\textit{Growth}) & \ding{56} & \ding{56} & \textbf{NA} & \textbf{NA} & \textbf{NA} & \textbf{NA} & \ding{56} & \textbf{NA} \\

        \textbf{STA-5} (\textit{Knowledge}) & \ding{56} & \textbf{$\boldsymbol{+}$} & \ding{56} & \textbf{$\boldsymbol{-}$} & \textbf{$\boldsymbol{-}$} & \ding{56} & \ding{56} & \ding{56}\\

        \textbf{STA-6} (\textit{Dormant}) & \ding{56} & \ding{56} & \ding{56} {\tiny (217)} & \ding{56} {\tiny (217)} & \ding{56} {\tiny (217)} & \ding{56} {\tiny (217)} & \ding{56} & \ding{56} {\tiny (217)}\\

        \textbf{STA-6} (\textit{Non Dormant}) & \ding{56} & \ding{56} & \ding{56} {\tiny (217)} & \ding{56} {\tiny (217)} & \ding{56} {\tiny (217)} & \ding{56} {\tiny (217)} & \ding{56} & \ding{56} {\tiny (217)}\\

        \textbf{STA-7} (\textit{Retention}) & \ding{56} & \textbf{$\boldsymbol{+}$} & \textbf{NA} & \textbf{NA} & \textbf{NA} & \textbf{NA} & \ding{56} & \textbf{NA}\\

        \textbf{STA-8} (\textit{Size})  & \textbf{$\boldsymbol{-}$} & \textbf{$\boldsymbol{+}$} & \textbf{$\boldsymbol{+}$} & \textbf{$\boldsymbol{-}$} & \textbf{$\boldsymbol{-}$} & \textbf{$\boldsymbol{-}$} & \ding{56} & \textbf{$\boldsymbol{-}$}\\

        \textbf{STA-9} (\textit{Turnover})  & \ding{56} & \textbf{$\boldsymbol{+}$} & \textbf{$\boldsymbol{+}$} {\tiny (212)} & \textbf{$\boldsymbol{-}$} {\tiny (212)} & \textbf{$\boldsymbol{-}$} {\tiny (212)} & \textbf{$\boldsymbol{-}$} {\tiny (212)} & \ding{56} & \textbf{$\boldsymbol{-}$} {\tiny (212)}\\ \midrule
        
        \textbf{TEC-1} (\textit{Dev. activity}) & \ding{56} & \textbf{$\boldsymbol{+}$} & \textbf{NA} & \textbf{NA} & \textbf{NA} & \textbf{NA} & \ding{56} & \textbf{NA} \\

        \textbf{TEC-2} (\textit{Efficiency}) & \ding{56} {\tiny (202)} & \ding{56} {\tiny (205)} & \textbf{$\boldsymbol{-}$} {\tiny (208)} & \textbf{$\boldsymbol{-}$} {\tiny (208)} & \textbf{$\boldsymbol{-}$} {\tiny (208)} & \textbf{$\boldsymbol{-}$} {\tiny (208)} & \ding{56}  {\tiny (206)} & \ding{56} {\tiny (208)} \\

        \textbf{TEC-3} (\textit{Non-code}) & \ding{56} & \textbf{$\boldsymbol{+}$} & \textbf{$\boldsymbol{+}$} & \textbf{$\boldsymbol{-}$} & \textbf{$\boldsymbol{-}$} & \textbf{$\boldsymbol{-}$} & \ding{56} & \textbf{$\boldsymbol{-}$} \\

        \textbf{TEC-4} (\textit{Overall dev.}) & \ding{56} & \ding{56} & \textbf{$\boldsymbol{+}$} {\tiny (214)} & \textbf{$\boldsymbol{-}$} {\tiny (214)} & \textbf{$\boldsymbol{-}$} {\tiny (214)} & \textbf{$\boldsymbol{-}$} {\tiny (214)} & \ding{56} & \textbf{$\boldsymbol{-}$} {\tiny (214)} \\
    
    \bottomrule
      
  \end{tabular}
  }

\end{table*}

\subsection{Communication}

\begin{figure}[th!]
  \centering
    \includegraphics[ trim=0cm 0cm 0cm 0cm, clip,
     width = .4 \textwidth
    ]{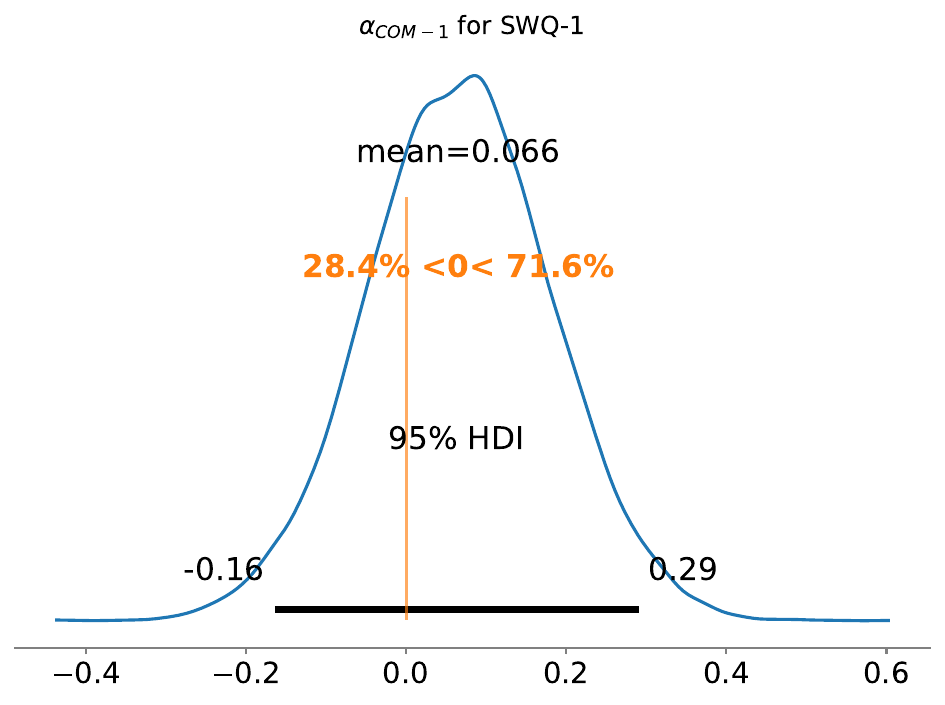}
    % \hspace{.01cm}
    \includegraphics[ trim=0cm 0cm 0cm 0cm, clip,
    width = .4 \textwidth
    ]{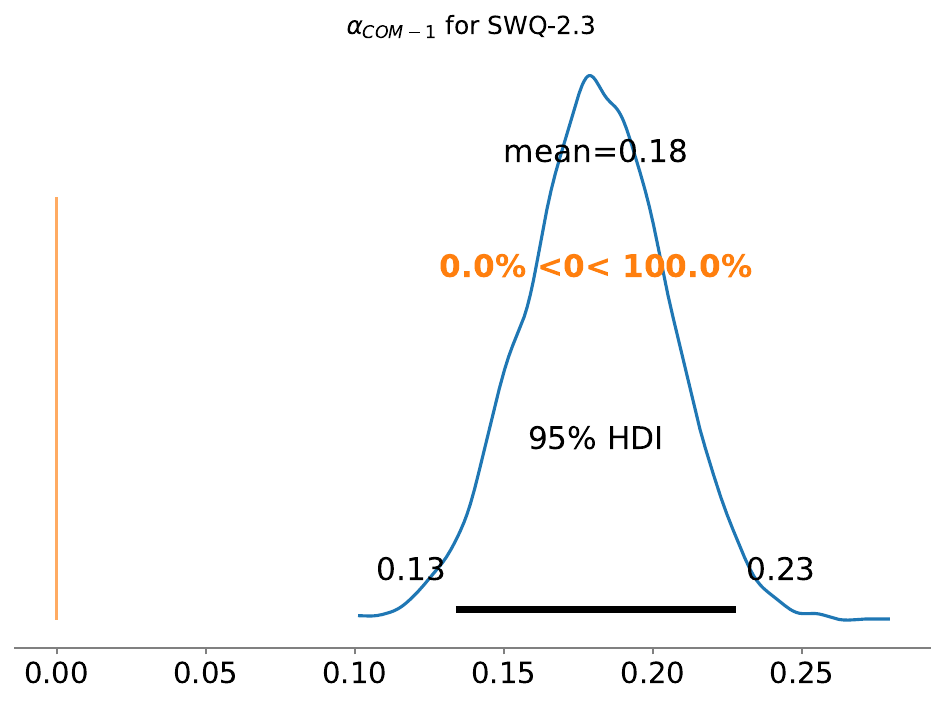}
    % \hspace{.01cm}
    \includegraphics[ trim=0cm 0cm 0cm 0cm, clip,
      width = .4 \textwidth
    ]{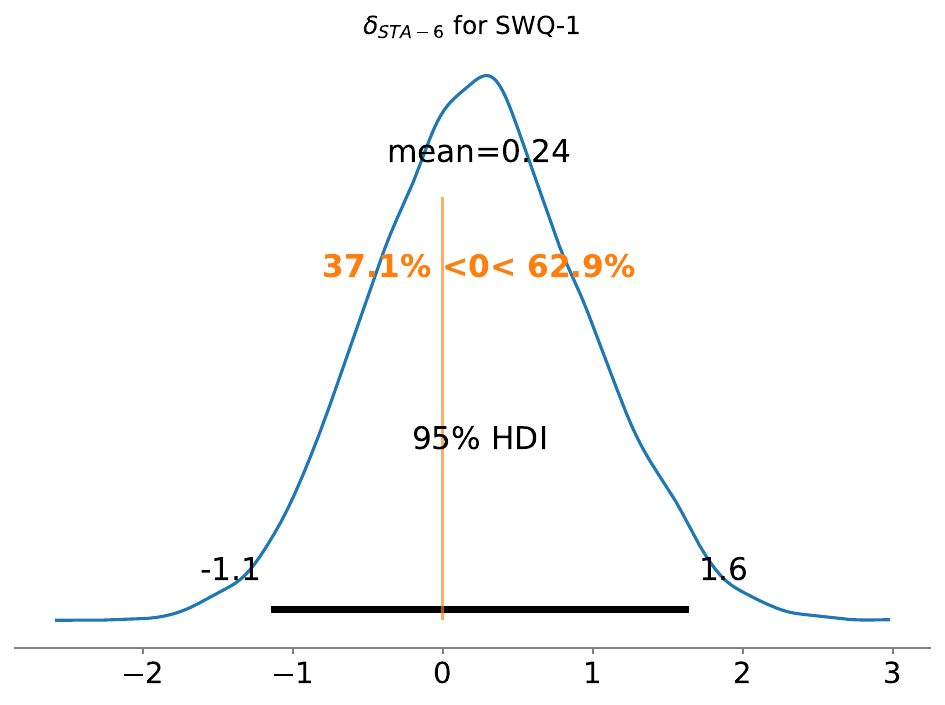}
    % \hspace{.01cm}
    \includegraphics[ trim=0cm 0cm 0cm 0cm, clip,
      width = .4 \textwidth
    ]{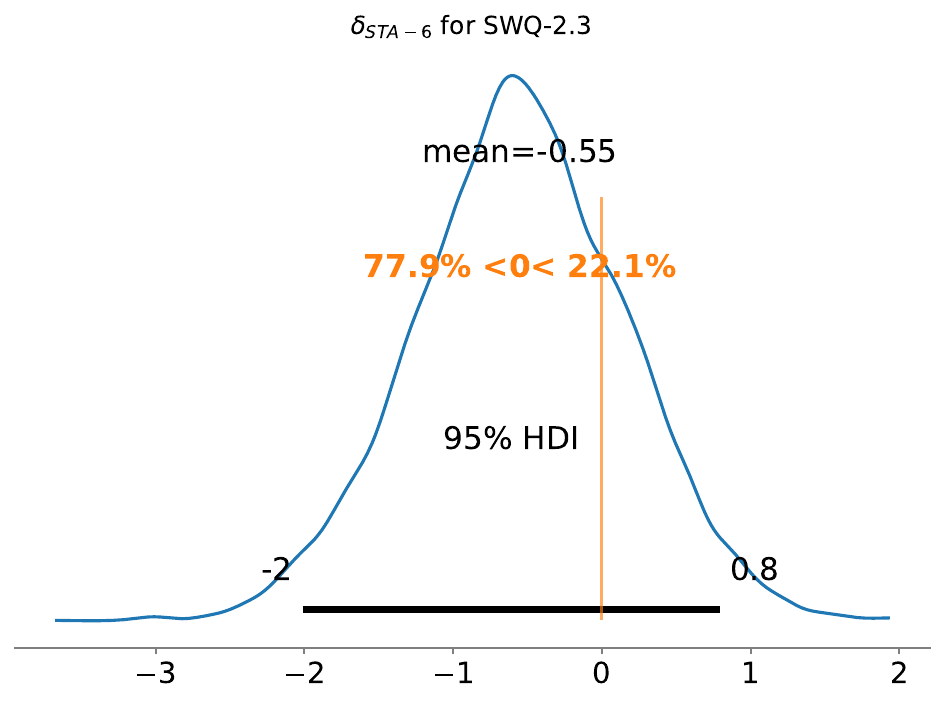}

    \caption{Posterior plots for the impact of response time (\textbf{COM-1}) on defect density (\textbf{SWQ-1}) (top-left) and very high risk complexity (\textbf{SWQ-2.3}) (top-right), and the impact of a project being dormant (\textbf{STA-6}) on defect density (\textbf{SWQ-1}) (bottom-left) and very high risk complexity (\textbf{SWQ-2.3}) (bottom-right).}%
    \label{fig:fig1}
\end{figure}

\paragraph{\textbf{H1a - Response time. }}
To illustrate our analysis, we provide a detailed interpretation of two posterior plots: \textbf{COM-1} (\emph{response time}) impact on \textbf{SWQ-1} (defect density) and \textbf{SWQ-2.3} (very high risk complexity) (Fig. \ref{fig:fig1}). We recall that our replication package contains the same details for the other cases in Table~\ref{tbl:result}---in total, the analysis of more than 100 sustainability/quality metric pairs.

For defect density \textbf{SWQ-1} (Fig.~\ref{fig:fig1}, left), the HDI of \textbf{COM-1} ranges from -0.16 to 0.29. Since 0 is part of the HDI, we conclude that there is a no statistical evidence to assert an impact of \textbf{COM-1} on this aspect of software quality (see our decision criteria, Sect. \ref{sec:methods}).
To better understand the information in the posterior distribution for analyses using the Gaussian model, we describe its precise interpretation for this case. Since we are using the Gaussian model, we have that \textbf{SWQ-1} was log-transformed, and \textbf{COM-1} was standardized. Thus, the interpretation of the results is as follows: one standard deviation change in \textbf{COM-1} entails a change for \textbf{SWQ-1} of $(e^{\alpha_{\mathit{COM-1}}}-1) \cdot 100$ percent. An increase of one standard deviation in \textbf{COM-1} is $31095852.87$ seconds ($\approx$ 1 year). The standard deviations for all metrics are computed using the collected project data. We refer interested readers to our replication package. Using the HDI for \textbf{COM-1} and the formula above, we obtain that \textbf{SWQ-1} can change from $-14\%$ to $33\%$. Thus, we conclude that \emph{an increase of approx. 1 year in response time entails a change from -0.09 decrease to 0.2 increase in defect density}.

For very high risk complexity \textbf{SWQ-2.3} (Fig.~\ref{fig:fig1}, right), the HDI of \textbf{COM-1} ranges from 0.13 to 0.23. Since 0 is not part of the HDI, we conclude that there is statistical evidence for a minor negative impact of \textbf{COM-1} on this aspect of code quality. We say negative impact because this result indicates that the higher the response time, the higher the number of functions with very high complexity. As mentioned above (Sect. \ref{sec:methods}), our decision criteria do take into account the impact strength, as impact strength depends on the metrics units, data transformations, and probabilistic model. Note that, as opposed to the example above, this analysis was carried out using a Poisson model. Therefore, we describe how to analyze the impact strength of this case as well. 
Since this analysis involves a Poisson regression model where \textbf{COM-1} was log-transformed, the interpretation is as follows: 10\% increase in \textbf{COM-1} is associated to an increase of $0.1 \cdot \alpha_{\mathit{COM-1}}$ change in number of functions with very high risk complexity (\textbf{SWQ-2.3}). In the projects we analyzed, an increase of 10\% in \textbf{COM-1} entails an increase of 45519321.12 seconds ($\approx$ 1.4 years) in response time. Thus, using the HDI for \textbf{COM-1} and the formula above, we can conclude that \emph{an increment of approx. 1.4 years in response time implies an increment between 0.013 and 0.023 in the number of functions with very high risk complexity.} These very low decrement values indicate that the strength in this decrement is minuscule. Yet all credible values in the posterior distribution of $\alpha_{\mathit{COM-1}}$ are less than 0. Therefore, this negative impact cannot confidently be discarded. 

Finally, we relate our results to the hypothesis we set forth in Sect~\ref{sec:hypotheses}. As illustrated in Tbl. \ref{tbl:result}, some \textbf{COM-1} excerpts have a positive impact on some code quality sub-metrics; however, this impact is not consistent. We observe a negative impact in most code quality sub-metrics, a positive impact only on \textbf{SWQ-2.2}, and no impact on \textbf{SWQ-1}, \textbf{SWQ-2.1}, and \textbf{SWQ-2.6}.
Therefore, we conclude that \textbf{H1a does not hold.} This may suggest that while certain aspects of code quality may benefit from faster community response times, overall, a longer response time correlates with challenges in maintaining high-quality and less complex code. This could also mean that rapid response times might not always be directly translated to higher software quality, and that the quality of responses and subsequent actions taken might be more important factors.

To conclude, while response time is an important aspect of community engagement, it is not a universal predictor of software quality across all metrics. This indicates the need to consider other aspects of community activity and software development practices when evaluating the determinants of software quality in FOSS communities.

\paragraph{\textbf{H1b - freq. of communication. }}

The analysis of this hypothesis explores the relationship between the frequency of communication within FOSS communities and various aspects of software quality metrics. Our findings present how increased communication activity, as measured by the number of comments and issues, influences software quality. A noteworthy increase in code coverage percentage changes (\textbf{SWQ-2.1}) indicates that more frequent communication could result in a slight improvement in this quality metric. Yet the impact on defect density (\textbf{SWQ-1}) and code quality metrics (\textbf{SWQ-2.3 ... SWQ-2.7} and \textbf{SWQ-2.7}) is negative (e.g., slight degradation in code duplication and complexity). Therefore, we conclude that \textbf{H1b does not hold.} These findings highlight the limitations of communication in directly influencing defect density, code duplication, and complexity. This may suggest that communication in the context of our sample, when it comes to quality, focuses more on code coverage improvement than on code optimization and complexity reduction.

\subsection{Popularity}

\paragraph{\textbf{H2 - Popularity. }}

\textbf{H2} investigates the impact of FOSS communities' popularity, as indicated by an increase in forks, stars, and watchers, on our selected software quality metrics. A notable finding is that the increased popularity resulted in an improvement in code coverage percentage (\textbf{SWQ-2.1}) and a slight decrease in the number of functions with medium complexity (\textbf{SWQ-2.2}). Conversely, our analysis shows no impact on defect density (\textbf{SWQ-1}) and a negative impact on other code quality metrics (\textbf{SWQ-2.3 ... SWQ-2.5}, and \textbf{SWQ-2.7}), suggesting that popularity might also result in increased code complexity and challenges in maintaining a lean and efficient codebase. Based on these mixed results, we conclude that \textbf{H2 does not hold.}

This may imply that popularity has a double-edged sword impact. While it can enhance certain quality aspects, like code coverage, it also brings challenges in other areas, such as defect density and code complexity. Our finding may suggest that while a project's visibility might attract resources, visibility, and attention, if only a few quality metrics improve, it does not necessarily translate into universal improvements across all dimensions of software quality.
\subsection{Stability}

\paragraph{\textbf{H3a - Age.}}

Our analysis shows that an increase in the age of FOSS communities in our sample results in a slight increase in defect density (\textbf{SWQ-1}) (i.e., a negative impact). Our analysis also shows a notable trend in metrics related to code complexity (\textbf{SWQ-2.3 ... SWQ-2.7}). An increase in the project's age tends to result in improvements in these code quality metrics. However, that is not the case for defect density (\textbf{SWQ-1}) and number of functions with medium complexity (\textbf{SWQ-2.2}). Hence, we conclude that \textbf{H3a does not hold.}

These findings may suggest that older projects are potentially accumulating more reported defects over time, which is a natural outcome of extended exposure and usage, hence the slight increase in defect density. These results also indicate that with age, FOSS communities become more mature at prioritizing and effectively managing code quality, addressing issues of complexity and maintainability. AFSI projects seem to evolve to become more focused on quality enhancement efforts as they stabilize and mature.

\paragraph{\textbf{H3b - Attrition.}}

\textbf{H3b} investigates the relationship between attrition in FOSS communities, as measured by the decrease in the number of commits, and its impact on our selected software quality metrics. As shown in Tbl. \ref{tbl:result}, an increase in attrition results in some code quality degradation, such as the number of functions with very high complexity (\textbf{SWQ-2.3}), the number of very large files (\textbf{SWQ-2.4}), the size of functions (\textbf{SWQ-2.5}), and lines of code for the most complex function (\textbf{SWQ-2.7}). The impact of attrition on our selected quality metrics is not consistent; our tests show no impact on \textbf{SWQ-1}, \textbf{SWQ-2.1}, and \textbf{SWQ-2.6}. Therefore, we conclude that \textbf{H3b does not hold.}

Even though the hypothesis does not hold, some of the results indicate that high attrition in FOSS communities is associated with a decrease in some aspects of code quality, particularly the adverse effects on code complexity and size. This finding may suggest that active and sustained contributor engagement is conducive to maintaining higher software quality practices.
 
\paragraph{\textbf{H3c - Forks.}}

\textbf{H3c} seeks to understand the effects of an increase in the number of forks within FOSS communities on our selected software quality metrics. One notable result is the positive impact on code coverage percentage (\textbf{SWQ-2.1}) and the number of functions with medium complexity (\textbf{SWQ-2.2}). Conversely, there are increases in the number of functions with very high complexity (\textbf{SWQ-2.3}), very large files (\textbf{SWQ-2.4}), and very large function size (\textbf{SWQ-2.5}) (i.e., negative impact as shown in Tbl. \ref{tbl:result}). Based on these mixed results, we conclude that \textbf{H3c does not hold.}

The positive impact on code coverage may suggest that projects are forked more frequently, possibly indicating higher interest or usage and a corresponding effort towards improving or maintaining rigorous testing. The negative impact on some code quality metrics also shows that forking might lead to diversification in project development paths, where some forks might introduce new features or new code that increase complexity and are not necessarily within quality expectations.
 
\paragraph{\textbf{H3d - Growth.}}

\textbf{H3d} aims to test the impact of growth in PR submissions on some software quality metrics. As shown in Tbl. \ref{tbl:result}, our tests indicate no impact (i.e., no statistical evidence to confirm a positive or negative impact) on \textbf{\textbf{SWQ-1}}, \textbf{\textbf{SWQ-2.1}}, and \textbf{\textbf{SWQ-2.6}}. In addition, our tests could not be computed for the remaining code quality metrics. Therefore, we conclude that \textbf{H3d does not hold.}

\paragraph{\textbf{H3e - Knowledge concentration.}} 

An interesting observation from our analysis is the results of \textbf{STA-5} (\emph{knowledge concentration}) or truck-factor. We find no statistical evidence for high knowledge concentration to increase or decrease on \textbf{SWQ-1}, \textbf{SWQ-2.2}, \textbf{SWQ-2.5}, \textbf{SWQ-2.6}, and \textbf{SWQ-2.7}, but a positive impact on \textbf{SWQ-2.1}, and a negative impact on \textbf{SWQ-2.3}, and \textbf{SWQ-2.4}. Therefore, we conclude that \textbf{H3e does not hold}. 

This finding may suggest that the concentration of knowledge within key individuals does not necessarily imply improvement or degradation in SWQ. It is plausible that other factors contribute to mitigating the risks associated with knowledge concentration, such as effective collaboration, documentation, and/or the dedicated efforts invested by maintainers to uphold SWQ standards. Also, in the context of ASFI, projects may tend to be relatively smaller compared to other FOSS communities. The impact of knowledge concentration may be less pronounced given the overall smaller scale of the development activities and the codebases, especially at the early stages of the project. Consequently, maintaining SWQ in these smaller projects may rely less on knowledge dispersion and more on the collective expertise and dedication of a few contributors and maintainers.

\paragraph{\textbf{H3f - Dormancy.}}

For both dormant and non-dormant projects, our tests show no impact on all our selected SWQ metrics (see Tbl. \ref{tbl:result}). Notably, we observed large HDIs in the posterior distribution of dormancy parameters for all our analyses, and they all include 0. For instance, in the bottom-row of figure~\ref{fig:fig1} we observe that the HDI for defect density goes from -1.1 to 1.6, and from -2 to 0.8 for high risk complexity. This is a common pattern for our analyses on dormancy. The width of these HDIs is approximately 2 to 4 times larger than for the other metrics. From the statistical point of view, this means that knowing whether a project is dormant or not does not convey much information regarding the quality metrics we considered. Hence, we conclude that \textbf{H3f is not supported.} 

\paragraph{\textbf{H3g - Retention.}}

Hypothesis \textbf{H3g} evaluates how changes in the retention of active contributors impact software quality metrics. A key observation is the positive in code coverage percentage (\textbf{SWQ-2.1}) associated with an increase in the number of active contributors (cf. Tbl.~\ref{tbl:result}). However, for the remaining, we could not either compute them or the tests indicate no impact. Thus, we conclude that \textbf{H3g is not supported.} These results may suggest that a stable and engaged contributor base leads to enhanced testing efforts.

\paragraph{\textbf{H3h - Size.}}

The analysis of \textbf{H3h} explores the relation between the size of FOSS communities, as indicated by the number of contributors engaging in commits, PRs, issues, or comments, and its impact on our selected software quality metrics. A noteworthy result is the positive impact on code coverage percentage (\textbf{SWQ-2.1}) with an increase in community size. Conversely, an increase in community size leads to a slight increase in defect density (\textbf{SWQ-1}). Moreover, larger communities show slight deterioration in code quality metrics (\textbf{SWQ-2.3} through \textbf{SWQ-2.5} and \textbf{SWQ-2.7}), except \textbf{SWQ-2.2} (number of functions with medium complexity). We conclude that our tests show an inconsistent pattern across selected software quality metrics. Therefore, we conclude that \textbf{H3h is not supported.}

These findings suggest that while an increase in FOSS community size brings some advantages to some aspects of software quality, notably code coverage and the number of functions with medium complexity, it also introduces challenges that can adversely affect other quality metrics. The deterioration in defect density could be attributed to higher defect reporting as the user and contributor base grow. However, the observed increase in complexity metrics with larger community sizes emphasizes the need for effective quality practices to leverage the benefits of a large contributor base.

\paragraph{\textbf{H3i - Turnover.}}

We proposed \textbf{H3i} to seek an understanding of the impact of turnover on our selected software quality metrics. Similar observation to last two sustainability metrics (i.e., \textbf{STA-8} and \textbf{STA-8}), despite high turnover, the impacts on code coverage (\textbf{SWQ-2.1}) and the number of functions with medium complexity (\textbf{SWQ-2.2}) remain positive. Our analysis also shows a negative impact in metrics associated with code complexity and size (\textbf{SWQ-2.3} through \textbf{SWQ-2.5}, \textbf{and SWQ-2.7}) with higher turnover rates. Therefore, we conclude that \textbf{the results do not uniformly support H3i as it was initially posited}. 

These results may suggest that despite high turnover, the remaining or new contributors are consistent in maintaining testing practices, demonstrated in code coverage. However, this trend is not mirrored across all code quality metrics related to code complexity and size. Given this pattern was observed in other sustainability indicators impacts, e.g., attrition and forks, then, a plausible explanation could be that code complexity and size characteristics are not communities priorities, but testing is.

\subsection{Technical activity}

\paragraph{\textbf{H4a - Contributors' dev. activity.}}

\textbf{H4a} seeks to examine the impact of development activity by contributors, not in maintainer roles, on various software quality metrics. A noteworthy observation is the positive impact on code coverage percentage (\textbf{SWQ-2.1}) when an increase in the number of non-maintainer contributors occurs. For defect density (\textbf{SWQ-1}) and code duplication percentage (\textbf{SWQ-2.6}), we observe no impact. For the remaining quality metrics, we could not carry out the analysis. Given the mixed outcomes observed across the tested software quality metrics, \textbf{the results do not uniformly support H4a as originally posited}.

\paragraph{\textbf{H4b - Efficiency.}}

\textbf{H4b} aims to test the relationship between the efficiency of FOSS communities, denoted by the time taken to merge or close PRs, and its impact on our selected software quality metrics. Our analysis shows that an increase in the time required to merge or close PRs results in a slight deterioration in some code quality metrics, particularly in the number of functions with medium and very high complexity (\textbf{SWQ-2.2} and \textbf{SWQ-2.3}), the number of very large files (\textbf{SWQ-2.4}), and the number of functions with very large function size (\textbf{SWQ-2.5}). For the remaining SWQ metrics, we observed no impact. Based on these results, we conclude that \textbf{H4b is not supported}. Given, other sustainability metrics close to efficiency in nature, i.e., response time and frequency of communication, have similar impact on code quality metrics, we can safely assume that the effect observed here is not a manifestation of sustainability but rather a result of communities prioritization of some aspect of SWQ (i.e., testing) over code simplification and optimization.

\paragraph{\textbf{H4c - Non-code contributions.}}

\textbf{H4c} seeks to understand the impact of non-code contributions, specifically commits to documentation files (txt and md), on several software quality metrics. A notable outcome of our analysis is the positive impact on code coverage percentage (\textbf{SWQ-2.1}) and the number of functions with medium complexity (\textbf{SWQ-2.2}). Further, we observed increases in complexity and size metrics (\textbf{SWQ-2.3} through \textbf{SWQ-2.5} and \textbf{SWQ-2.7}) (i.e., negative impact on code quality as shown in Tbl. \ref{tbl:result}) with a rise in documentation contributions. Given the mixed impacts observed, \textbf{H4c} does not find uniform support across all software quality metrics. Therefore, we conclude that \textbf{H4c is not supported.}

The positive impact on code coverage shows the potential benefits of non-code contributions in enhancing one aspect of software quality. One explanation could be that better documented code is easier to test and maintain, possibly contributing to more thorough testing practices and thus higher code coverage. Admittedly, the relationship between documentation contributions and software quality may be complex to test using only quantitative measures. While these contributions form an essential part of the software development ecosystem, influencing quality not only through direct effects on understandability and maintainability, how they help contributors may require a more nuanced investigation.

\paragraph{\textbf{H4d - Overall dev. activity.}}

\textbf{H4d} posits a negative relationship between the decline in FOSS communities' overall development activity and software quality. We tested whether decreased coding activity is associated with lower software quality. Our analysis show that an increase in commits related to coding files exacerbate code complexity and size metrics (\textbf{SWQ-2.3 ... SWQ-2.5}, and \textbf{SWQ-2.7}), except for number of functions with medium complexity (\textbf{SWQ-2.2}), showing a minor improvement. For \textbf{SWQ-1}, \textbf{SWQ-2.1}, and \textbf{SWQ-2.6}, we observed no impact. Hence, we conclude that \textbf{H4d is not supported.} This could be interpreted as a sign that simply more coding activities, in the context of projects in our sample, lead to more complex and potentially less maintainable code.

\paragraph{\textbf{Summary: }}To conclude, our investigation into the impact of FOSS community sustainability on software quality in the context of ASFI projects reveals a complex relationship. Our analysis across multiple sustainability themes---communication, popularity, stability, and technical activity---shows a mixed picture. Certain sustainability indicators, such as increased community response time, age, and community size, exhibit both negative and positive impacts on different software quality metrics. Notably, while some aspects, like code coverage, improve with heightened frequency in communication, others, such as code complexity and size, experience a degradation in quality. This complexity has become more evident in the role of non-code contributions, indicating that documentation may enhance specific aspects of quality but require more nuanced exploration to fully understand their impact.

Our findings also challenge the simplistic view of the relationship between quality and FOSS sustainability, which underscores the importance of adopting a holistic and context-aware approach to understanding this relationship. In addition, sustainability strategies may not always be uniformly effective in improving all aspects of software quality. For example, we can note that our sustainability indicators consistently exhibit either a positive impact or no impact on code coverage. This suggests that improvements in certain sustainability indicators enhance specific quality metrics while leaving others unaffected.

Alternatively, communities in our sample may emphasize their quality assurance on testing and less on code quality. This observation is in line with the broader trend in software engineering teams. As defects are more visible and may impact the community's reputation, testing and code coverage are often prioritized as immediate and measurable indicators of software quality. In addition, FOSS projects attract contributors with varying levels of expertise and familiarity with the codebase. Hence, emphasizing code coverage as a quality assurance practice could serve as a practical strategy to assure basic quality standards. More focus on testing than on reducing code complexity or size might reflect the pragmatic attitude of these communities. Taking into consideration the constraints under which these communities operate, balancing the need for quality assurance with the diverse contributions characteristic of open source development can be a challenge.

In response to our RQ: \textbf{How does the decline in the sustainability of FOSS communities affect their software quality?}, we found a multifaceted relationship between the sustainability of FOSS communities and the quality of the software they produce. Our findings suggest that the decline in the sustainability of FOSS communities affects their software quality in complex and varied ways. The decrease in certain sustainability indicators can be associated with both improvements and deteriorations in software quality metrics. For instance, while an increase in community response frequency and size positively impacts code coverage, indicating an improvement in one aspect of software quality, these same factors can lead to increased code complexity and size, suggesting a degradation in other quality aspects.
\section{Discussion and Implications}\label{sec:discussion}

We begin this section by contextualizing our findings within the framework of extant literature (summarized in Sect. \ref{sec:related}), after which we follow by a discussion of potential interpretations. From these discussions, we will formulate actionable recommendations tailored for projects within ASFI. 

\paragraph{\textbf{Contexualization with extant work: }} \textit{St{\u{a}}nciulescu, et al.} used ``graduation'' and ``retirement'' as outcomes equating ``sustainable'' and ``unsustainable'' \cite{stuanciulescu2022code}. They found that higher number of files with more than 1 KLOC negatively impacts graduation \cite{stuanciulescu2022code}. Other metrics influencing the graduation outcome are: increases in lines of code, increases in major and minor contributors, increases in features commits, increases in corrective commits, presence of medium complexity functions, and presence of very large functions \cite{stuanciulescu2022code}. They also found that code coverage has no significant impact on project's sustainability \cite{stuanciulescu2022code}. In contrast, we opted to use metrics drawn from existing literature to indicate signs of sustainability or its decline. Our approach assumes that the decisions made for graduation and retirement do not necessarily reflect the project sustainability. This assumption is grounded in the fact that ASFI projects graduation and retirement decisions are influenced by several factors which do not always fit with what has been reported in the literature as sustainability indicators. Importantly, we do not claim that our approach is superior to the one adopted by St{\u{a}}nciulescu, et al., but rather, we view it as a legitimate and complementary avenue of investigation. In addition, while St{\u{a}}nciulescu, et al. focused on code quality \cite{stuanciulescu2022code}, we extended the outcome variable to cover both code quality and defects density representing an extended coverage of software quality attributes (see Sect. \ref{sec:swq}).

\textit{Ghapanchi}'s work correlates certain project features such as the rates of defect fixing and feature addition with sustainability, implying direct impact on project longevity \cite{ghapanchi2015predicting}. This work's approach aligns well with the traditional view of functionality-focused assessment of quality and sustainability. In contrast, our work used a broad array of themes and their indicators to investigate the impact of sustainability on SWQ. While Ghapanchi's study found positive correlations between defect reduction, feature addition, and sustainability \cite{ghapanchi2015predicting}, we found no conclusive evidence linking our adopted sustainability indicators to SWQ. This may imply that SWQ in FOSS communities maybe influenced with more complex factors, social in nature, such as motivation for quality \cite{alami2019does}, and other norms inclined towards promoting better quality \cite{alami2020foss,alami2021pull}. For example, Alami et al. found that in the case of the ROS community, quality assurance practices are influenced by factors peculiar to the community itself \cite{alami2018influencers}. For instance, in the context of robotic software development (the ROS community software), the community prioritize innovation over SWQ \cite{alami2018influencers}. Despite the ROS community effort to establish and promote quality assurance practices, contributors deem experimenting and innovating more important than adhering to quality expectations \cite{alami2018influencers}. The case of ROS underscores that SWQ is sometimes seen through the lens of the community own values and objectives. In the case of ASFI, projects could be more focused on their credibility, especially prior graduation to demonstrate the merit to graduate. This focus may become ingrained in the project culture post graduation. These community-driven objectives may imply that understanding SWQ should be contextual, considering the unique cultural, motivational, and normative factors that define each community.

\textit{Foucault et al.}'s study found a negative impact of turnover on module' quality \cite{foucault2015impact}. However, our study did not find statistical evidence of turnover impact on defect density, code coverage, and inconsistent outcome on code quality metrics. This may suggest that the relationship between FOSS sustainability indicators, like turnover, and SWQ is more intricate than what we assumed. Potentially, other variables moderate the relationship, such as code review, decision-making in pull request (PR) process, or the codebase itself. Even though inconsistent, notably, certain code quality metrics, especially very high risk complexity, number of very large files, and Very Large Function Size Count are negatively impacted even when projects showing signs of sustainability. Do{\u{g}}an and and T{\"u}z{\"u}n study on code review smells shows that when the code change is large and complex, developers are unwilling to review and when they do the review, then it is ineffective \cite{dougan2022towards}. This observation may suggest that FOSS environments, development practices, such as code review and decision-making in pull requests, also play a role in shaping software quality. It appears that sustainability indicators may not fully capture an impact on SWQ, unless considered alongside other factors specific to development and quality assurance practices.

\textit{Wang et al.}'s study found that when elite developers contribute more to organizational and supporting tasks, some SWQ metrics such as bug fixing rate degrades \cite{wang2020unveiling}. Our study found mixed impact of knowledge concentration of certain SWQ metrics (i.e., either no impact, positive (code coverage), or negative on some code quality metrics). This discrepancy may imply that the impact of knowledge concentration on software quality is context-dependent, not necesseraly universal, and may vary across communities. In the case of ASFI project, they tend to be small in size (especially retired ones) \cite{stuanciulescu2022code}. Increase concentration among elite contributors may not make a difference. Because in small projects, they tend to be the key contributors. This indicates that the relationship between individual contributors' roles and SWQ is complex and might not be generalizable. The unique characteristics of each FOSS community also have a role.

\textit{We may conclude}, from this discussion so far, that the relationship between FOSS sustainability and software quality is non-linear and complex, and maybe a disputed proxy for FOSS sustainability. Yet, in the context of ASFI, the results of our study have several potential interpretations and implications. As we found that code quality metrics decline even in a state of sustainability. This result could be rooted in several behaviors, and conditions: awareness and perception of code quality requirements, and codebase growth.

\paragraph{\textbf{Awareness and perception of code quality.}} Projects and their contributors may not be cognizant of established code quality practices and the metrics used for the evaluation of code quality. They may also hold their own distinct interpretations of what constitutes code quality. B{\"o}rstler et al. studied software developers' perception of code quality \cite{borstler2023developers}. They found that developers prioritize the attributes: readability, structure, comprehensibility, maintainability, correctness, documentation, dynamic behavior, testability, correctness, maintainability, analysis, generic and miscellaneous. Their results, notably, assert that developers have affinity for code structure, readability and documentation. However, they reported that more experienced developers and those with quality assurances (QA) responsibilities tend to focus on correctness. Less experienced developer focus on code documentation \cite{borstler2023developers}. 

Even though, most of these expectations resonates with traditional software engineering view on code quality, a noticeable observation is the focus on readability, and structure which may render some of our metrics obsolete. For example, a complex and large function if it is well ``structured'' and ``readable,'' then certain metrics like very high risk complexity, number of very large files, very large function size count may become irrelevant in the eyes of contributors. The emphasis on readability and structure by developers, as opposed to more traditional metrics, suggests a possible shift in how code quality is assessed in practice. This could challenge the relevance of certain conventional metrics, we used in our study. In projects, where a well-structured and readable code is prioritized as code quality, then other parameters such as complexity and size are less relevant. Some metrics, cyclomatic complexity in particular, received criticism \cite{shepperd1994critique}, yet they remain popular \cite{menzies2006data}. A recent literature survey on the topic by Klima et al. reveals that these metrics are highly adopted in the software engineering community \cite{klima2022selected}.

These projects usually attract contributors with varying levels of experience and expertise in developing software and they may not always have a thorough understanding of code quality best practices and how to apply them. Alami et al. studied the decision-making mechanisms in PR process \cite{alami2021pull,alami2020foss}. They found that some communities strategically adopt mentoring of new contributors to help them meet established expectations for quality. This strategy is also part of a broader agenda to promote a ``lenient'' culture where contributor are treated fairly, promoting retention \cite{alami2021pull}. Fagerholm et al.'s study corroborates these findings \cite{fagerholm2014role}. They suggest that contributors receiving support through mentoring are more ``active'' and can contribute more efficiently compared to those who do not receive mentoring \cite{fagerholm2014role}. However, Steinmacher et al. suggest that mentors should receive proper training and support for such program to be efficient \cite{steinmacher2021being}.

ASFI may consider integrating awareness for code quality through mentoring. This could be supplemented by knowledge sharing artifacts for code quality (for example, code quality check list, and training materials, etc.). ASFI may consider including code quality in its mentorship programs. The aim of the mentoring activities is to guide projects and their contributors in adhering to code quality practices. Mentors should provide guidance on coding standards, best practices for code quality, and promote the importance of maintaining high-quality code.

\paragraph{\textbf{Codebase growth. }} The codebase becomes more extensive and complex, as projects grow, become sustainable, and their software evolves. This growth may come at the expense of code quality. It is a well-established claim that as project's code grow in size and become more complex, it's design quality degrades and technical debt grows \cite{mannan2018towards,ahmed2015empirical}. Our findings resonate with this claim. Our tests of project's size show negative impact on most code quality metrics, except medium risk complexity and code duplication percentage. As project expand and grow in size, the intricacies of the codebase also may increase, potentially leading to challenges in maintaining quality, which could be partially due to code readability and understandability \cite{mannan2018towards,ahmed2015empirical}. In addition, code smells makes new contributions and maintenance difficult \cite{fowler1997refactoring}. Changes to the codebase may become complex with unintended consequences; code edits may lead to more bugs and poor code \cite{khomh2012exploratory}.

Managing and contributing to complex and large codebase brings challenges especially to newcomers. Larger projects may face more challenges maintaining code quality due to the sheer volume of contributions and code changes. It may also become more challenging to enforce consistent code quality standards. Alami et al. found that some FOSS communities have consistent styles of enforcing quality expectations \cite{alami2021pull}. Some communities have a ``transactional'' style, quality expectations are applied rigidly and indiscriminately during PR reviews, other have more ``lenient'' approach and prefer to mentor newcomers, and those who unfamiliar with the community standards and codebase. The communities studied also have strong adherence to software engineering and quality assurance practices, e.g., ``PR atomicity'', peer review, and documented best practices for coding \cite{alami2021pull}. Other studies highlighted similar approaches, e.g., \cite{anderson2020assessing}. For example, Anderson et al. reported that Network Automation communities are more ``inclusive'', a deliberate strategy by the projects to attract new contributors and remain sustainable, yet quality is a priority \cite{anderson2020assessing}. This work show the importance of adherence to quality assurance practices and fostering a culture that prioritizes code quality, through inclusively, mentoring, and fostering good adherence to software engineering practices.

ASFI may consider including code quality in its mentorship programs. The aim of the mentoring activities is to guide projects and their contributors in balancing between new features and code quality. Project's mentors should provide guidance on coding standards, best practices for code quality, and promote the importance of maintaining high-quality code. Fostering a culture that prioritizes code quality, should also target developers. Projects should encourage developers to take ownership of the code's quality.

\section{Limitations \& Threats to validity}\label{sec:validity}

\paragraph{\textbf{Limitations: }} 
Our analysis does not delineate between ``graduated'' and ``retired'' projects in contrast to Stuanciulescu et al.'s (2022) work. This deviation arises because we opted to use a concise adaptation of the Lin{\aa}ker et al. \cite{linaaker2022characterize} framework for evaluating sustainability. ASFI uses a set of subjective criteria and a community-driven process (e.g., voting, board approval, etc.), and these may not necessarily be aligned with evaluating sustainability; e.g., a project may have exhibited promising sustainability signs but is still retired because of negative voting or a lack of approval from the board for other reasons unrelated to its sustainability. Such analysis could illuminate findings on how sustainability manifests in ``graduated'' versus ``retired'' projects.

Our adaptation of the Lin{\aa}ker et al. \cite{linaaker2022characterize} framework has excluded themes and indicators not available in repository data such as ``finance'' and ``culture'' and/or not well-acknowledged in the literature. However, the inherent constraints of our method have made the full adoption of the framework with its 15 themes and 107 indicators unfeasible. Sourcing some of these indicators may require an exhaustive investigation across 217 projects and may not be applicable in the context of ASFI. It is worth noting that the implementation of the framework can be iterative. While we prioritized the most significant and available indicators, we also ensured that our study captured a representative and meaningful aspect of the sustainability concept.

Klima et al. conducted a comprehensive review of code quality metrics in the context of IoT systems \cite{klima2022selected}. In this study, sourced form code quality metrics literature, they consolidated several metrics, and revised them based in their relevance for IoT systems. They also suggested further metrics to capture attributes relevant to the source code of IoT systems. This review suggest nine major code quality categories: size, redundancy, complexity, coupling, unit test coverage and effectiveness, cohesion, code readability, security, and code heterogeneity \cite{klima2022selected}. Even though, we did not adopt all proposed metrics (44 metrics), our coverage of software quality metrics is comprehensive, covering key metrics in Klima et al.'s study and supersedes similar studies in FOSS sustainability.

\paragraph{\textbf{Internal validity: }} 
Data availability and quality. As briefly mentioned in the methods section (Sect. \ref{sec:methods}), we encountered data availability and quality issues during the mining activity. Not all ASFI projects have well-maintained repositories. For example, not all projects have JIRA issues labeled; this has constrained our ability to compute \textbf{SWQ-1} for some projects. This may have introduced bias in the selection process. Nevertheless, our sample size remains large and contains projects with diverse scopes, maturity levels, and sustainability levels. This diversity enhances the validity of our conclusions.

Prior distributions and decision criteria. We selected standard non-informative priors; i.e., priors that weigh all parameter values equally (no matter how unlikely they are). This is a common choice when no prior information is available. However, these priors may over-represent the probability of parameter values that are implausible. Similarly, our decision criteria are not sensitive to the strength of the impact we measure. This choice was made to have consistent criteria for all our analyses. However, cases where we report positive/negative impact may effectively be so small that are insignificant in practice.

Exclusion of process metrics. We sought to understand tangible aspects of software products, and strategically emphasized product-focused metrics to assess software quality, and not process metrics (e.g., release-duration, the number of commits made to a file, and the number of developers who changed a file) \cite{rahman2013and}. This focus allowed us a standardized and objective measurement of software quality that is broadly applicable to various FOSS communities. Future work could use the combination of product and process-focused metrics to evaluate the relationship between FOSS sustainability and software quality.

The difference in data availability across projects of varying ages is also an internal validity threat. This issue is inherent to the variability in the volume of commits, issues, and other project activities, which are lower for younger projects compared to their older counterparts. These disparities in data may have inadvertently influenced the analysis. The analytical process may have yielded less reliable insights for younger projects due to the sparsity of data points. However, our sample has a selection of projects with wide range of ages. This choice is to deliberately capture a broad spectrum of FOSS project lifecycle stages. 

Our analysis considers single predictor models. That is, we have studied how a single sustainability metric impacts a single quality metric. However, it is possible that combinations of sustainability metrics have an impact while each of them individually do not. To study these effects, it is necessary to build multiple predictor models. Unfortunately, these models introduce complications in the interpretability of the results---as results are interpreted conditioned on a subset of predictors (e.g.,~\cite{mcelreath2020}). On the contrary, our models have a simple and direct interpretation (see Sect.~\ref{sec:methods}). This work aims at building a solid and easy-to-interpret foundation in the understanding of the interplay between sustainability and quality metrics. For this reason, we have considered single predictor models in this work; and we leave as future work the study of the impact of combination of sustainability metrics on software quality.

\paragraph{\textbf{External validity: }} Heterogeneity of FOSS communities. FOSS communities outside the context of ASFI vary largely, from small, niche communities to communities with widely used products. Understandably, sustainability may take different shapes in these instances. Hence, we do not claim the generalization of our findings beyond our sample. In addition, ASFI projects may not be representative of the broader FOSS communities. While acknowledging the heterogeneity of FOSS communities, our findings provide valuable insights to the ASFI community and can serve as a reference point for future research.

Exclusion criteria. We excluded projects without PRs in GitHub or Jira, empty issue trackers, etc. This choice may have led to a selection bias. The final sample could be over-representative of certain types of projects. Our selection criteria favor projects with well-structured software development infrastructure, which may affect the generalizability of our findings. 

Considering other factors alongside sustainability indicators. We strictly focused on sustainability metrics, which may not be the sole influencers of software quality. Acknowledging other factors, such as community-driven priorities, could have further strengthened the external validity. Our focus allowed us a clear and detailed analysis of several sustainability factors, which allowed us a deeper understanding of the sustainability effect on software quality.

\section{Conclusion}\label{sec:conclusion}

In conclusion, we find that the relationship between an OSS community's sustainability and quality of the community's software outputs is multifaceted and complex. We considered 17 sustainability metrics across the themes of communication, popularity, stability, and technical activity, and note that a decline among these can be associated with both improvements and deterioration in eight software quality metrics leveraged. For example, while an increase in community response frequency and size positively impacts code coverage, indicating an improvement in one aspect of software quality, these same factors can lead to increased code complexity and size, suggesting a degradation in other quality aspects.

These results show that the sustainability of FOSS, as measured by these specific indicators, is not necessarily linked to code quality practices in the context of ASFI projects. Interestingly, older projects seem to show mature code quality practices, as reflected in some of the metrics we used. We conclude that our results may indicate that growth resulting in a more complex and extensive codebase, coupled with a potential lack of awareness of code quality standards reflected in our metrics, may explain the deterioration in some aspects of code quality.

Practitioners are recommended to not blindly evaluate the sustainability of a FOSS community based on the quality of its software outputs. Rather, a more comprehensive picture is needed, including extent and quality to which testing and quality assurance is performed by the community, and how this practice is documented and transferred, e.g., via mentoring. 

Researchers should look further into the relationships between sustainability indicators and its impact on software quality beyond the ASFI projects investigated in this study. Future work is further encouraged to go beyond software quality metrics to expand investigations to also include testing and quality assurance practices, and its relation to sustainability markers. Qualitative research is also recommended to complement presented findings, as they can shed further light on how sustainability is perceived from the concerned communities, and how they consider the relation to software quality.

\section{Statements and Declarations}

\paragraph{Data Availability Statements.}\label{sec:replication}

\noindent We made this study's data and other artifacts available \href{https://doi.org/10.5281/zenodo.10629606}{here.}\footnote{\url{https://doi.org/10.5281/zenodo.10629606}}

\paragraph{Conflict of Interest.} The authors declared that they have no conflict of interest.

\section*{Acknowledgments}\label{sec:acks}

No external funding was received to carry out this study. All resources utilized in the study were provided by the authors' affiliated institutions.

\bibliographystyle{spmpsci}
\bibliography{references}

\begin{thebibliography}{10}
\providecommand{\url}[1]{{#1}}
\providecommand{\urlprefix}{URL }
\expandafter\ifx\csname urlstyle\endcsname\relax
  \providecommand{\doi}[1]{DOI~\discretionary{}{}{}#1}\else
  \providecommand{\doi}{DOI~\discretionary{}{}{}\begingroup
  \urlstyle{rm}\Url}\fi

\bibitem{ahmed2015empirical}
Ahmed, I., Mannan, U.A., Gopinath, R., Jensen, C.: An empirical study of design
  degradation: How software projects get worse over time.
\newblock In: 2015 ACM/IEEE International Symposium on Empirical Software
  Engineering and Measurement (ESEM), pp. 1--10. IEEE (2015)

\bibitem{alami2020foss}
Alami, A., Cohn, M.L., W{\k{a}}isowski, A.: How do foss communities decide to
  accept pull requests?
\newblock In: Proceedings of the 24th International Conference on Evaluation
  and Assessment in Software Engineering, pp. 220--229 (2020)

\bibitem{alami2019does}
Alami, A., Cohn, M.L., W{\k{a}}sowski, A.: Why does code review work for open
  source software communities?
\newblock In: 2019 IEEE/ACM 41st International Conference on Software
  Engineering (ICSE), pp. 1073--1083. IEEE (2019)

\bibitem{alami2018influencers}
Alami, A., Dittrich, Y., W{\k{a}}sowski, A.: Influencers of quality assurance
  in an open source community.
\newblock In: Proceedings of the 11th International Workshop on Cooperative and
  Human Aspects of Software Engineering, pp. 61--68 (2018)

\bibitem{alami2022scrum}
Alami, A., Krancher, O.: How scrum adds value to achieving software quality?
\newblock Empirical Software Engineering \textbf{27}(7), 165 (2022)

\bibitem{alami2021pull}
Alami, A., Pardo, R., Cohn, M.L., W{\k{a}}sowski, A.: Pull request governance
  in open source communities.
\newblock IEEE Transactions on Software Engineering \textbf{48}(12), 4838--4856
  (2021)

\bibitem{anderson2020assessing}
Anderson, J., Steinmacher, I., Rodeghero, P.: Assessing the characteristics of
  foss contributions in network automation projects.
\newblock In: 2020 IEEE International Conference on Software Maintenance and
  Evolution (ICSME), pp. 324--335. IEEE (2020)

\bibitem{avelino2016novel}
Avelino, G., Passos, L., Hora, A., Valente, M.T.: A novel approach for
  estimating truck factors.
\newblock In: 2016 IEEE 24th International Conference on Program Comprehension
  (ICPC), pp. 1--10. IEEE (2016)

\bibitem{borges2016understanding}
Borges, H., Hora, A., Valente, M.T.: Understanding the factors that impact the
  popularity of github repositories.
\newblock In: 2016 IEEE international conference on software maintenance and
  evolution (ICSME), pp. 334--344. IEEE (2016)

\bibitem{borstler2023developers}
B{\"o}rstler, J., Bennin, K.E., Hooshangi, S., Jeuring, J., Keuning, H.,
  Kleiner, C., MacKellar, B., Duran, R., St{\"o}rrle, H., Toll, D., et~al.:
  Developers talking about code quality.
\newblock Empirical Software Engineering \textbf{28}(6), 128 (2023)

\bibitem{champion2021underproduction}
Champion, K., Hill, B.M.: Underproduction: An approach for measuring risk in
  open source software.
\newblock In: 2021 IEEE International Conference on Software Analysis,
  Evolution and Reengineering (SANER), pp. 388--399. IEEE (2021)

\bibitem{chengalur2010sustainability}
Chengalur-Smith, I., Sidorova, A., Daniel, S.L.: Sustainability of free/libre
  open source projects: A longitudinal study.
\newblock Journal of the Association for Information Systems \textbf{11}(11), 5
  (2010)

\bibitem{coelho2017modern}
Coelho, J., Valente, M.T.: Why modern open source projects fail.
\newblock In: Proceedings of the 2017 11th Joint meeting on foundations of
  software engineering, pp. 186--196 (2017)

\bibitem{crowston2005social}
Crowston, K., Howison, J.: The social structure of free and open source
  software development.
\newblock First Monday  (2005)

\bibitem{crowston2008free}
Crowston, K., Wei, K., Howison, J., Wiggins, A.: Free/libre open-source
  software development: What we know and what we do not know.
\newblock ACM Computing Surveys (CSUR) \textbf{44}(2), 1--35 (2008)

\bibitem{dougan2022towards}
Do{\u{g}}an, E., T{\"u}z{\"u}n, E.: Towards a taxonomy of code review smells.
\newblock Information and Software Technology \textbf{142}, 106737 (2022)

\bibitem{fagerholm2014role}
Fagerholm, F., Guinea, A.S., M{\"u}nch, J., Borenstein, J.: The role of
  mentoring and project characteristics for onboarding in open source software
  projects.
\newblock In: Proceedings of the 8th ACM/IEEE international symposium on
  empirical software engineering and measurement, pp. 1--10 (2014)

\bibitem{fang2009understanding}
Fang, Y., Neufeld, D.: Understanding sustained participation in open source
  software projects.
\newblock Journal of Management Information Systems \textbf{25}(4), 9--50
  (2009)

\bibitem{foucault2015impact}
Foucault, M., Palyart, M., Blanc, X., Murphy, G.C., Falleri, J.R.: Impact of
  developer turnover on quality in open-source software.
\newblock In: Proceedings of the 2015 10th joint meeting on foundations of
  software engineering, pp. 829--841

\bibitem{Log4j}
Foundation, T.A.S.: Log4j – apache log4j™ 2.
\newblock \url{https://logging.apache.org/log4j/2.x/}.
\newblock (Accessed on 04/18/2023)

\bibitem{fowler1997refactoring}
Fowler, M., Beck, K.: Refactoring: Improving the design of existing code.
\newblock In: 11th European Conference. Jyv{\"a}skyl{\"a}, Finland (1997)

\bibitem{furia2022applying}
Furia, C.A., Torkar, R., Feldt, R.: Applying bayesian analysis guidelines to
  empirical software engineering data: The case of programming languages and
  code quality.
\newblock ACM Transactions on Software Engineering and Methodology (TOSEM)
  \textbf{31}(3), 1--38 (2022)

\bibitem{ghapanchi2015predicting}
Ghapanchi, A.H.: Predicting software future sustainability: A longitudinal
  perspective.
\newblock Information Systems \textbf{49}, 40--51 (2015)

\bibitem{guizani2021long}
Guizani, M., Chatterjee, A., Trinkenreich, B., May, M.E., Noa-Guevara, G.J.,
  Russell, L.J., Cuevas~Zambrano, G.G., Izquierdo-Cortazar, D., Steinmacher,
  I., Gerosa, M.A., et~al.: The long road ahead: Ongoing challenges in
  contributing to large oss organizations and what to do.
\newblock Proceedings of the ACM on Human-Computer Interaction
  \textbf{5}(CSCW2), 1--30 (2021)

\bibitem{han2019characterization}
Han, J., Deng, S., Xia, X., Wang, D., Yin, J.: Characterization and prediction
  of popular projects on github.
\newblock In: 2019 IEEE 43rd annual computer software and applications
  conference (COMPSAC), vol.~1, pp. 21--26. IEEE (2019)

\bibitem{nuts}
Hoffman, M.D., Gelman, A., et~al.: The no-u-turn sampler: adaptively setting
  path lengths in hamiltonian monte carlo.
\newblock J. Mach. Learn. Res. \textbf{15}(1), 1593--1623 (2014)

\bibitem{iso2011}
ISO/IEC: Iso/iec 25010:2011(en) systems and software engineering — systems
  and software quality requirements and evaluation (square) — system and
  software quality models (2011).
\newblock
  \urlprefix\url{https://www.iso.org/obp/ui/#iso:std:iso-iec:25010:ed-1:v1:en}

\bibitem{jansen2014measuring}
Jansen, S.: Measuring the health of open source software ecosystems: Beyond the
  scope of project health.
\newblock Information and Software Technology \textbf{56}(11), 1508--1519
  (2014)

\bibitem{jensen2007role}
Jensen, C., Scacchi, W.: Role migration and advancement processes in ossd
  projects: A comparative case study.
\newblock In: 29th International Conference on Software Engineering (ICSE'07),
  pp. 364--374. IEEE (2007)

\bibitem{jiang2019metrics}
Jiang, S., Cao, J., Prasad, M.: The metrics to evaluate the health status of
  oss projects based on factor analysis.
\newblock In: Computer Supported Cooperative Work and Social Computing: 14th
  CCF Conference, ChineseCSCW 2019, Kunming, China, August 16--18, 2019,
  Revised Selected Papers 14, pp. 723--737. Springer (2019)

\bibitem{khomh2012faster}
Khomh, F., Dhaliwal, T., Zou, Y., Adams, B.: Do faster releases improve
  software quality? an empirical case study of mozilla firefox.
\newblock In: 2012 9th IEEE working conference on mining software repositories
  (MSR), pp. 179--188. IEEE (2012)

\bibitem{khomh2012exploratory}
Khomh, F., Penta, M.D., Gu{\'e}h{\'e}neuc, Y.G., Antoniol, G.: An exploratory
  study of the impact of antipatterns on class change-and fault-proneness.
\newblock Empirical Software Engineering \textbf{17}, 243--275 (2012)

\bibitem{kitchenham1996software}
Kitchenham, B., Pfleeger, S.L.: Software quality: the elusive target [special
  issues section].
\newblock IEEE software \textbf{13}(1), 12--21 (1996)

\bibitem{klima2022selected}
Klima, M., Bures, M., Frajtak, K., Rechtberger, V., Trnka, M., Bellekens, X.,
  Cerny, T., Ahmed, B.S.: Selected code-quality characteristics and metrics for
  internet of things systems.
\newblock IEEE Access \textbf{10}, 46144--46161 (2022)

\bibitem{kruschke2018bayesian}
Kruschke, J.K., Liddell, T.M.: The bayesian new statistics: Hypothesis testing,
  estimation, meta-analysis, and power analysis from a bayesian perspective.
\newblock Psychonomic bulletin \& review \textbf{25}, 178--206 (2018)

\bibitem{linaaker2022characterize}
Lin{\aa}ker, J., Papatheocharous, E., Olsson, T.: How to characterize the
  health of an open source software project? a snowball literature review of an
  emerging practice.
\newblock In: Proceedings of the 18th International Symposium on Open
  Collaboration, pp. 1--12 (2022)

\bibitem{manikas2013reviewing}
Manikas, K., Hansen, K.M.: Reviewing the health of software ecosystems--a
  conceptual framework proposal.
\newblock In: Proceedings of the 5th international workshop on software
  ecosystems (IWSECO), pp. 33--44. Citeseer (2013)

\bibitem{mannan2018towards}
Mannan, U.A., Ahmed, I., Sarma, A.: Towards understanding code readability and
  its impact on design quality.
\newblock In: Proceedings of the 4th ACM SIGSOFT International Workshop on NLP
  for Software Engineering, pp. 18--21 (2018)

\bibitem{mcelreath2020}
McElreath, R.: Statistical Rethinking: A Bayesian Course with Examples in R and
  STAN, 2 edn.
\newblock CRC Press (2020)

\bibitem{menzies2006data}
Menzies, T., Greenwald, J., Frank, A.: Data mining static code attributes to
  learn defect predictors.
\newblock IEEE transactions on software engineering \textbf{33}(1), 2--13
  (2006)

\bibitem{midha2012factors}
Midha, V., Palvia, P.: Factors affecting the success of open source software.
\newblock Journal of Systems and Software \textbf{85}(4), 895--905 (2012)

\bibitem{mockus2002two}
Mockus, A., Fielding, R.T., Herbsleb, J.D.: Two case studies of open source
  software development: Apache and mozilla.
\newblock ACM Transactions on Software Engineering and Methodology (TOSEM)
  \textbf{11}(3), 309--346 (2002)

\bibitem{pymc}
Oriol, A.P., Virgile, A., Colin, C., Larry, D., J., F.C., Maxim, K., Ravin, K.,
  Jupeng, L., C., L.C., A., M.O., Michael, O., Ricardo, V., Thomas, W., Robert,
  Z.: Pymc: A modern and comprehensive probabilistic programming framework in
  python.
\newblock {PeerJ} Computer Science \textbf{9}, e1516 (2023).
\newblock \doi{10.7717/peerj-cs.1516}

\bibitem{osman2021health}
Osman, K., Baysal, O.: Health is wealth: Evaluating the health of the bitcoin
  ecosystem in github.
\newblock In: 2021 IEEE/ACM 4th International Workshop on Software Health in
  Projects, Ecosystems and Communities (SoHeal), pp. 1--8. IEEE (2021)

\bibitem{rahman2013and}
Rahman, F., Devanbu, P.: How, and why, process metrics are better.
\newblock In: 2013 35th International Conference on Software Engineering
  (ICSE), pp. 432--441. IEEE (2013)

\bibitem{ray2014large}
Ray, B., Posnett, D., Filkov, V., Devanbu, P.: A large scale study of
  programming languages and code quality in github.
\newblock In: Proceedings of the 22nd ACM SIGSOFT international symposium on
  foundations of software engineering, pp. 155--165 (2014)

\bibitem{shah2013software}
Shah, S.M.A., Morisio, M., Torchiano, M.: Software defect density variants: A
  proposal.
\newblock In: 2013 4th International Workshop on Emerging Trends in Software
  Metrics (WETSoM), pp. 56--61. IEEE (2013)

\bibitem{shaikh2019selecting}
Shaikh, M., Levina, N.: Selecting an open innovation community as an alliance
  partner: Looking for healthy communities and ecosystems.
\newblock Research Policy \textbf{48}(8), 103766 (2019)

\bibitem{shepperd1994critique}
Shepperd, M., Ince, D.C.: A critique of three metrics.
\newblock Journal of systems and software \textbf{26}(3), 197--210 (1994)

\bibitem{stuanciulescu2022code}
St{\u{a}}nciulescu, È., Yin, L., Filkov, V.: Code, quality, and process metrics
  in graduated and retired asfi projects.
\newblock In: Proceedings of the 30th ACM Joint European Software Engineering
  Conference and Symposium on the Foundations of Software Engineering, pp.
  495--506 (2022)

\bibitem{steinmacher2021being}
Steinmacher, I., Balali, S., Trinkenreich, B., Guizani, M., Izquierdo-Cortazar,
  D., Cuevas~Zambrano, G.G., Gerosa, M.A., Sarma, A.: Being a mentor in open
  source projects.
\newblock Journal of Internet Services and Applications \textbf{12}(1), 1--33
  (2021)

\bibitem{steinmacher2019overcoming}
Steinmacher, I., Gerosa, M., Conte, T.U., Redmiles, D.F.: Overcoming social
  barriers when contributing to open source software projects.
\newblock Computer Supported Cooperative Work (CSCW) \textbf{28}, 247--290
  (2019)

\bibitem{valiev2018ecosystem}
Valiev, M., Vasilescu, B., Herbsleb, J.: Ecosystem-level determinants of
  sustained activity in open-source projects: A case study of the pypi
  ecosystem.
\newblock In: Proceedings of the 2018 26th ACM Joint Meeting on European
  Software Engineering Conference and Symposium on the Foundations of Software
  Engineering, pp. 644--655 (2018)

\bibitem{van2017health}
Van~Vulpen, P., Menkveld, A., Jansen, S.: Health measurement of data-scarce
  software ecosystems: a case study of apple's researchkit.
\newblock In: Software Business: 8th International Conference, ICSOB 2017,
  Essen, Germany, June 12-13, 2017, Proceedings 8, pp. 131--145. Springer
  (2017)

\bibitem{vasilescu2015quality}
Vasilescu, B., Yu, Y., Wang, H., Devanbu, P., Filkov, V.: Quality and
  productivity outcomes relating to continuous integration in github.
\newblock In: Proceedings of the 2015 10th joint meeting on foundations of
  software engineering, pp. 805--816 (2015)

\bibitem{walden2020impact}
Walden, J.: The impact of a major security event on an open source project: The
  case of openssl.
\newblock In: Proceedings of the 17th International Conference on Mining
  Software Repositories, pp. 409--419 (2020)

\bibitem{wang2012survival}
Wang, J.: Survival factors for free open source software projects: A
  multi-stage perspective.
\newblock European Management Journal \textbf{30}(4), 352--371 (2012)

\bibitem{wang2020unveiling}
Wang, Z., Feng, Y., Wang, Y., Jones, J.A., Redmiles, D.: Unveiling elite
  developers' activities in open source projects.
\newblock ACM Transactions on Software Engineering and Methodology (TOSEM)
  \textbf{29}(3), 1--35 (2020)

\bibitem{xia2022predicting}
Xia, T., Fu, W., Shu, R., Agrawal, R., Menzies, T.: Predicting health
  indicators for open source projects (using hyperparameter optimization).
\newblock Empirical Software Engineering \textbf{27}(6), 122 (2022)

\bibitem{yin2021sustainability}
Yin, L., Chen, Z., Xuan, Q., Filkov, V.: Sustainability forecasting for apache
  incubator projects.
\newblock In: Proceedings of the 29th ACM Joint Meeting on European Software
  Engineering Conference and Symposium on the Foundations of Software
  Engineering, pp. 1056--1067 (2021)

\bibitem{zhou2012make}
Zhou, M., Mockus, A.: What make long term contributors: Willingness and
  opportunity in oss community.
\newblock In: 2012 34th International Conference on Software Engineering
  (ICSE), pp. 518--528. IEEE (2012)

\end{thebibliography}

\end{document}